\documentclass[12pt]{iopart}
\usepackage{graphicx}

\begin{document}

\title{Thermalization of quark-gluon matter with the elastic scattering of
$gqq$, $gq\bar{q}$ and $g\bar{q}\bar{q}$}

\author{Xiao-Ming Xu$^1$ and Li-Sha Xu$^2$}

\address{$^1$ Department of Physics, Shanghai University, Baoshan,
Shanghai 200444, China}
\address{$^2$ Department of Information Engineering, China Jiliang University, 
Hangzhou 310018, China}
\ead{xmxu@xmxucao.sina.net}
\begin{abstract}
The elastic scattering of $gqq$, $gq\bar{q}$ and $g\bar{q}\bar{q}$ and the
thermalization of quark-gluon matter are studied. According to Feynman diagrams
at the tree level, squared amplitudes for the elastic $gqq$ scattering and the
elastic $gq\bar{q}$ scattering are derived in perturbative QCD. Transport
equations including the 
squared amplitudes for the elastic $gqq$, $gq\bar{q}$ and 
$g\bar{q}\bar{q}$ scattering are established. Corresponding to anisotropic
gluon and quark distributions created in central Au-Au collisions at RHIC,
solutions of the transport equations show that thermalization time of quark
matter can be shortened by the elastic $gqq$, $gq\bar{q}$ and $g\bar{q}\bar{q}$
scattering. 
\end{abstract}

\pacs{24.85.+p;12.38.Mh;12.38.Bx;25.75.Nq}

\section{Introduction}
Scattering takes place in quark-gluon matter that is created in high-energy 
heavy-ion collisions. If quark-gluon matter has a low number density, two-body
scattering substantially affects the evolution of quark-gluon matter. The 
two-body scattering includes the 2-to-2 scattering and the 2-to-3 scattering
\cite{shuryak,geiger,bdmtw,wong,ndmg,sm,xg}. If the number density is high, 
the three-body scattering becomes
important. This has been shown by the elastic gluon-gluon-gluon scattering in 
gluon matter which has a number density of the order of
19 ${\rm fm}^{-3}$ that is reached 
in central Au-Au collisions at the Relativistic Heavy Ion Collider (RHIC)
\cite{xmxu3}. One effect derived from the elastic gluon-gluon-gluon
scattering is on the rapid thermalization of gluon matter. The elastic
3-to-3 scattering of a heavy quark contributes significantly to 
the heavy quark momentum degradation in the quark-gluon plasma \cite{liuko}.
Since at the Large Hadron Collider (LHC) Pb-Pb collisions will be carried out
to produce quark-gluon matter which has a higher number 
density than matter reached at RHIC, the elastic 3-to-3 scattering is more
involved. Therefore, we must study the elastic 3-to-3 scattering and its 
effects. Due to the complication of the study \cite{xmxu4}, in this work 
we are restricted to the elastic gluon-quark-quark scattering, the elastic
gluon-quark-antiquark scattering and the elastic gluon-antiquark-antiquark 
scattering, and apply the scattering to thermalization of quark-gluon matter.

We display Feynman diagrams at the tree level for the elastic gluon-quark-quark
scattering in Section 2 and the elastic gluon-quark-antiquark scattering in 
Section 3. In Section 4, as two examples, we show squared amplitudes 
for a diagram of the elastic $gqq$ scattering and for a diagram
of the elastic $gq\bar q$ scattering, respectively. We present transport 
equations that include the elastic scattering of $gqq$, $gq\bar q$ and 
$g\bar{q}\bar q$. In Section 5 numerical solutions of the transport equations
and relevant discussions are given. The last section contains the summary.

\section{Elastic gluon-quark-quark scattering}
Some Feynman diagrams for the elastic gluon-quark-quark scattering are shown 
in Figs. 1-4. The wiggly lines and solid lines stand for gluons and quarks, 
respectively. The other Feynman diagrams are derived from the diagrams in Figs.
1-4 as follows.

The six diagrams in Fig. 1 lead to six new diagrams by moving the external 
gluons from the left quark line to the right quark line, i.e. by
moving the initial gluon from an initial
(final) quark to another initial (final) quark and moving the final gluon in
the same way. If the two final quarks are identical, the exchange of the final
quarks in the above twelve diagrams leads to twelve more diagrams. We can thus 
derive 18 diagrams from the six diagrams in Fig. 1. 

The four diagrams in Fig. 2
lead to four new diagrams by moving the initial gluon from the initial
(final) state of a quark to the final (initial) state of the same quark and 
moving the final gluon in the same way. If the two final quarks are identical,
the exchange of the final quarks in the above eight diagrams leads to eight
more diagrams. We can thus derive 12 diagrams from the four diagrams in Fig. 2.

For the elastic scattering of one gluon and two identical quarks, we need to
take into account 40 diagrams that do not contain any triple-gluon vertex.
Each of the 40 diagrams contains one gluon propagator and four 
gluon-quark vertices. The 40 diagrams form the first class of processes
for the elastic $gqq$ scattering.

Every diagram in Fig. 3 contains one triple-gluon vertex.
We can derive 18 diagrams from the six diagrams. Each of the diagrams
${\rm D}_{-\rm M}$ and ${\rm D}_{+\rm M}$ contains one gluon propagator
between the triple-gluon vertex and the left quark. The diagrams 
${\rm D}_{-\rm M}$ and ${\rm D}_{+\rm M}$ lead to two new diagrams by letting 
the gluon propagating from the triple-gluon vertex to the right quark. The
diagrams ${\rm D}_{\rm GUML}$, ${\rm D}_{\rm GUMH}$, ${\rm D}_{\rm GDMH}$ and 
${\rm D}_{\rm GDML}$ lead to four new diagrams by moving the initial or final 
gluon from the left quark to the right
quark. If the two final quarks are identical, the exchange of the final quarks 
in the above twelve diagrams generates twelve more diagrams.

While one triple-gluon vertex is involved in the elastic scattering of one
gluon and two identical quarks, twenty-four diagrams need to be taken into
consideration. Each of the 24 diagrams contains two gluon propagators and three
gluon-quark vertices. The 24 diagrams form the second class of processes
for the elastic $gqq$ scattering.

The four diagrams in Fig. 4 are characterized by two triple-gluon vertices or 
one four-gluon vertex, and give rise to four more diagrams by the exchange of
the two final quarks if the quarks are identical. The eight diagrams contain
three or two gluon propagators and two
gluon-quark vertices. The 8 diagrams form the third class of 
processes for the elastic $gqq$ scattering.

We have arrived at the three classes of processes represented by the 72 
diagrams 
for the elastic scattering of one gluon and two identical quarks. If the two
quarks do not possess the same flavor, the processes with the exchange of final
quarks do not happen and thirty-six diagrams are needed. To ensure gauge 
invariance we include elastic ghost-quark-quark scattering for which some 
Feynman diagrams are shown in Fig. 5. Ghosts are indicated by the dashed lines.
Each of the diagrams ${\rm D}_{-\rm MFP}$ and ${\rm D}_{+\rm MFP}$ contains one
gluon propagator between the ghost line and the left quark line. If the gluon
propagator is connected to the right quark line, two new diagrams are generated
by
the diagrams ${\rm D}_{-\rm MFP}$ and ${\rm D}_{+\rm MFP}$. In total, we need
7 diagrams for distinguishable quarks and 14 diagrams for indistinguishable  
quarks.

\section{Elastic gluon-quark-antiquark scattering}
Since quark-antiquark annihilation may happen in elastic gluon-quark-antiquark
scattering, the scattering involves more Feynman diagrams than the elastic 
gluon-quark-quark scattering. We only display some diagrams in Figs.
6-12, but the other diagrams can de derived as follows.

In each of the six diagrams in Fig. 6 the
initial and final gluons are attached to the quark line. While the initial
and final gluons are attached to the antiquark line, six new diagrams are
produced. 
The four diagrams in Fig. 7 lead to four new diagrams by moving an external
gluon from the final quark to the initial quark and moving another
external gluon from the initial (final) antiquark to the final (initial)
antiquark. Hence, there are twenty diagrams of which any contains 
no self-coupling of gluons and no quark-antiquark annihilation.

The external gluons in Fig. 8 are attached to one or two of the quark lines.
Similarly, the initial and final gluons can be attached to one or two of 
the antiquark lines and six new diagrams are thus created. 
In Fig. 9 the four diagrams lead to four new diagrams by moving one external
gluon from the final quark to the initial quark and moving another 
external gluon from the initial (final) antiquark to the final (initial) 
antiquark. Therefore, twenty diagrams exist in the case of quark-antiquark
annihilation and no self-coupling of gluons.

In total, from Figs. 6-9 we have 40 diagrams of which each contains 
no triple-gluon coupling and no four-gluon coupling. The 40 diagrams form the 
first class of processes for the elastic $gq\bar q$ scattering.

Each of the diagrams ${\rm E}_{-\rm M}$ and ${\rm E}_{+\rm M}$ in Fig. 10
contains one gluon propagator between the triple-gluon vertex and the quark 
line. The two diagrams
generate two new diagrams while the gluon propagator is connected to the 
antiquark line. The diagrams ${\rm E}_{\rm GUML}$, ${\rm E}_{\rm GUMH}$,
${\rm E}_{\rm GDMH}$ and ${\rm E}_{\rm GDML}$ generate four new diagrams while 
the external gluon irrelevant to the triple-gluon vertex is moved to the 
antiquark line. Therefore, twelve diagrams correspond to the case in which 
every diagram has one triple-gluon vertex and quark-antiquark annihilation and 
creation do not take place.

The diagram ${\rm E}_{\rm QD}$ (${\rm E}_{\rm QU}$) in Fig. 11 contains one
gluon propagator between the triple-gluon vertex and the initial (final) quark.
The diagrams ${\rm E}_{\rm QD}$ and ${\rm E}_{\rm QU}$ give rise to two new 
diagrams while the gluon propagates between the triple-gluon vertex and the 
initial or final antiquark. In the diagram ${\rm E}_{\rm QDE}$ the initial
quark radiates a gluon that creates a quark-antiquark pair.
The diagram ${\rm E}_{\rm QDE}$ generates one new diagram by moving the gluon 
to the initial antiquark. In the diagram 
${\rm E}_{\rm QUE}$ the initial quark-antiquark pair annihilates into a gluon
that is absorbed by the final quark. The
diagram ${\rm E}_{\rm QUE}$ generates one new diagram by moving the gluon to 
the final antiquark. Four new diagrams are given by the diagrams
${\rm E}_{\rm GUMLA}$, ${\rm E}_{\rm GUMHA}$,
${\rm E}_{\rm GDMHA}$ and ${\rm E}_{\rm GDMLA}$ by moving the external gluon 
irrelevant to the triple-gluon vertex from a quark to an antiquark. Finally,
sixteen diagrams correspond to the case in which every diagram has one 
triple-gluon vertex and possesses quark-antiquark annihilation and creation.

In total, from Figs. 10 and 11 we have 28 diagrams of which each
contains one triple-gluon coupling. The 28 diagrams 
form the second class of processes for the elastic $gq\bar q$ scattering.

Any of the diagrams in Fig. 12 has two triple-gluon couplings or one
four-gluon coupling. The 8 diagrams form the third class of processes for the
elastic $gq\bar q$ scattering.

We have arrived at the three classes of processes shown by the 76 diagrams. 
If the quark-antiquark annihilation may happen, all the 76 diagrams
must be considered. If the 
annihilation does not occur, thirty-six diagrams are needed.
To satisfy gauge invariance we include elastic ghost-quark-antiquark 
scattering. Some Feynman diagrams for the scattering are plotted in Figs. 13 
and 14. In Fig. 13 each of the diagrams ${\rm E}_{-\rm MFP}$, 
${\rm E}_{+\rm MFP}$, ${\rm E}_{\rm QDFP}$ and ${\rm E}_{\rm QUFP}$
has one gluon propagator between the ghost line and the quark line. If the 
gluon propagator is between the ghost line and the antiquark line, four new 
diagrams are obtained from the diagrams ${\rm E}_{-\rm MFP}$, 
${\rm E}_{+\rm MFP}$, ${\rm E}_{\rm QDFP}$ and ${\rm E}_{\rm QUFP}$. In the
diagram ${\rm E}_{\rm QDEFP}$ the gluon radiated from the initial quark breaks
into a quark-antiquark pair. If the gluon is radiated from the initial
antiquark, one new
diagram is generated from the diagram ${\rm E}_{\rm QDEFP}$. In the diagram 
${\rm E}_{\rm QUEFP}$ the gluon
from the annihilation of a quark-antiquark pair is absorbed by the final quark.
If the gluon is absorbed by the final antiquark, a new diagram is generated
from the diagram ${\rm E}_{\rm QUEFP}$. Finally, we have 6 diagrams derived
from the six diagrams in Fig. 13 and together with the six diagrams in
Fig. 14 we need to consider 18 diagrams for the elastic ghost-quark-antiquark
scattering.

\section{Transport equations}
We establish transport equations for quark-gluon matter which consists of 
gluons, quarks and antiquarks with up and down flavors.
Denote the gluon distribution function by $f_{gi}$ where $i$ labels the 
$i$th gluon in scattering. Let the distribution functions for the 
up quark, the down quark, the up antiquark and the down antiquark be
$f_{ui}$, $f_{di}$, $f_{\bar {u}i}$ and $f_{\bar {d}i}$, respectively, and we
assume that they are identical,
\begin{equation}
f_{ui}=f_{di}=f_{\bar {u}i}=f_{\bar {d}i}=f_{qi},
\end{equation}
where $i$ labels the $i$th quark or antiquark in 
scattering. With elastic 2-to-2 scattering and elastic 3-to-3 
scattering, the transport equation for gluons is
\begin{eqnarray}
\fl
\frac {\partial f_{g1}}{\partial t}
+ \vec {\rm v}_1 \cdot \vec {\nabla}_{\vec {r}} f_{g1}
         \nonumber    \\
\fl
= -\frac {1}{2E_1} \int \frac {d^3p_2}{(2\pi)^32E_2}
\frac {d^3p_3}{(2\pi)^32E_3} \frac {d^3p_4}{(2\pi)^32E_4}
(2\pi)^4 \delta^4(p_1+p_2-p_3-p_4)
         \nonumber    \\
\fl
~~~ \times \left\{ \frac {g_G}{2} \mid {\cal M}_{gg \to gg} \mid^2
[f_{g1}f_{g2}(1+f_{g3})(1+f_{g4})-f_{g3}f_{g4}(1+f_{g1})(1+f_{g2})]  \right.
         \nonumber    \\
\fl
~~~ + g_Q ( \mid {\cal M}_{gu \to gu} \mid^2
+ \mid {\cal M}_{gd \to gd} \mid^2
+ \mid {\cal M}_{g\bar {u} \to g\bar {u}} \mid^2
+ \mid {\cal M}_{g\bar {d} \to g\bar {d}} \mid^2 )
         \nonumber    \\
\fl
~~~ \left.
\times [f_{g1}f_{q2}(1+f_{g3})(1-f_{q4})-f_{g3}f_{q4}(1+f_{g1})(1-f_{q2})]
    \right\}
         \nonumber    \\
\fl
~~~ -\frac {1}{2E_1} \int \frac {d^3p_2}{(2\pi)^32E_2}
\frac {d^3p_3}{(2\pi)^32E_3} \frac {d^3p_4}{(2\pi)^32E_4}
\frac {d^3p_5}{(2\pi)^32E_5} \frac {d^3p_6}{(2\pi)^32E_6}
         \nonumber    \\
\fl
~~~ \times (2\pi)^4 \delta^4(p_1+p_2+p_3-p_4-p_5-p_6) 
\left\{ \frac {g_G^2}{12} \mid {\cal M}_{ggg \to ggg} \mid^2  \right.
         \nonumber    \\
\fl
~~~ \times [f_{g1}f_{g2}f_{g3}(1+f_{g4})(1+f_{g5})(1+f_{g6})-
f_{g4}f_{g5}f_{g6}(1+f_{g1})(1+f_{g2})(1+f_{g3})]
         \nonumber    \\
\fl
~~~ + \frac {g_Gg_Q}{2} ( \mid {\cal M}_{ggu \to ggu} \mid^2
+\mid {\cal M}_{ggd \to ggd} \mid^2
+\mid {\cal M}_{gg\bar {u} \to gg\bar {u}} \mid^2 
+\mid {\cal M}_{gg\bar {d} \to gg\bar {d}} \mid^2 )
         \nonumber    \\
\fl
~~~ \times [f_{g1}f_{g2}f_{q3}(1+f_{g4})(1+f_{g5})(1-f_{q6})
-f_{g4}f_{g5}f_{q6}(1+f_{g1})(1+f_{g2})(1-f_{q3})]
         \nonumber    \\
\fl
~~~ + g_Q^2 [\frac {1}{4} \mid {\cal M}_{guu \to guu} \mid^2
+\frac {1}{2} ( \mid {\cal M}_{gud \to gud} \mid^2
              + \mid {\cal M}_{gdu \to gdu} \mid^2 )
+\frac {1}{4} \mid {\cal M}_{gdd \to gdd} \mid^2
         \nonumber    \\
\fl
~~~ + \mid {\cal M}_{gu\bar {u} \to gu\bar {u}} \mid^2
    + \mid {\cal M}_{gu\bar {d} \to gu\bar {d}} \mid^2
    + \mid {\cal M}_{gd\bar {u} \to gd\bar {u}} \mid^2
    + \mid {\cal M}_{gd\bar {d} \to gd\bar {d}} \mid^2
         \nonumber         \\
\fl
~~~ +\frac {1}{4} \mid {\cal M}_{g\bar {u}\bar {u}
                             \to g\bar {u}\bar {u}} \mid^2
    +\frac {1}{2} ( \mid {\cal M}_{g\bar {u}\bar {d}
                               \to g\bar {u}\bar {d}} \mid^2 
    + \mid {\cal M}_{g\bar {d}\bar {u} \to g\bar {d}\bar {u}} \mid^2 )
+ \frac {1}{4} \mid {\cal M}_{g\bar {d}\bar {d} \to g\bar {d}\bar {d}} \mid^2 ]
         \nonumber    \\
\fl
~~~ \left. \times [f_{g1}f_{q2}f_{q3}(1+f_{g4})(1-f_{q5})(1-f_{q6})
           -f_{g4}f_{q5}f_{q6}(1+f_{g1})(1-f_{q2})(1-f_{q3})] \right\} ,     
         \nonumber    \\
\end{eqnarray}
and the transport equation for up quarks is
\begin{eqnarray}
\fl
\frac {\partial f_{q1}}{\partial t}
+ \vec {\rm v}_1 \cdot \vec {\nabla}_{\vec {r}} f_{q1}
         \nonumber    \\
\fl
= -\frac {1}{2E_1} \int \frac {d^3p_2}{(2\pi)^32E_2}
\frac {d^3p_3}{(2\pi)^32E_3} \frac {d^3p_4}{(2\pi)^32E_4}
(2\pi)^4 \delta^4(p_1+p_2-p_3-p_4)
         \nonumber    \\
\fl
~~~ \times \left\{ g_G \mid {\cal M}_{ug \to ug} \mid^2
[f_{q1}f_{g2}(1-f_{q3})(1+f_{g4})-f_{q3}f_{g4}(1-f_{q1})(1+f_{g2})]  \right.
         \nonumber    \\
\fl
~~~ + g_Q (\frac {1}{2} \mid {\cal M}_{uu \to uu} \mid^2
+ \mid {\cal M}_{ud \to ud} \mid^2 
+ \mid {\cal M}_{u\bar {u} \to u\bar {u}} \mid^2
+ \mid {\cal M}_{u\bar {d} \to u\bar {d}} \mid^2 )
         \nonumber    \\
\fl
~~~ \times \left. [f_{q1}f_{q2}(1-f_{q3})(1-f_{q4})
-f_{q3}f_{q4}(1-f_{q1})(1-f_{q2})]   \right\}
         \nonumber    \\
\fl
~~~ -\frac {1}{2E_1} \int \frac {d^3p_2}{(2\pi)^32E_2}
\frac {d^3p_3}{(2\pi)^32E_3} \frac {d^3p_4}{(2\pi)^32E_4}
\frac {d^3p_5}{(2\pi)^32E_5} \frac {d^3p_6}{(2\pi)^32E_6}
         \nonumber    \\
\fl
~~~ \times (2\pi)^4 \delta^4(p_1+p_2+p_3-p_4-p_5-p_6)
\left\{ \frac {g_G^2}{4} \mid {\cal M}_{ugg \to ugg} \mid^2  \right.
         \nonumber    \\
\fl
~~~ \times [f_{q1}f_{g2}f_{g3}(1-f_{q4})(1+f_{g5})(1+f_{g6})
-f_{q4}f_{g5}f_{g6}(1-f_{q1})(1+f_{g2})(1+f_{g3})]
          \nonumber     \\
\fl
~~~ +g_Qg_G ( \frac {1}{2} \mid {\cal M}_{uug \to uug} \mid^2
+\mid {\cal M}_{udg \to udg} \mid^2
+\mid {\cal M}_{u\bar {u}g \to u\bar {u}g} \mid^2
+\mid {\cal M}_{u\bar {d}g \to u\bar {d}g} \mid^2 )
         \nonumber     \\
\fl
~~~ \times [f_{q1}f_{q2}f_{g3}(1-f_{q4})(1-f_{q5})(1+f_{g6})
-f_{q4}f_{q5}f_{g6}(1-f_{q1})(1-f_{q2})(1+f_{g3})]
          \nonumber     \\
\fl
~~~ + g_Q^2 [\frac {1}{12} \mid {\cal M}_{uuu \to uuu} \mid^2
+\frac {1}{4} ( \mid {\cal M}_{uud \to uud} \mid^2
              + \mid {\cal M}_{udu \to udu} \mid^2 )
+\frac {1}{4} \mid {\cal M}_{udd \to udd} \mid^2
         \nonumber    \\
\fl
~~~ +\frac {1}{2} \mid {\cal M}_{uu\bar {u} \to uu\bar {u}} \mid^2
    +\frac {1}{2} \mid {\cal M}_{uu\bar {d} \to uu\bar {d}} \mid^2
            + \mid {\cal M}_{ud\bar {u} \to ud\bar {u}} \mid^2
            + \mid {\cal M}_{ud\bar {d} \to ud\bar {d}} \mid^2
         \nonumber         \\
\fl
~~~ +\frac {1}{4} \mid {\cal M}_{u\bar {u}\bar {u}
                             \to u\bar {u}\bar {u}} \mid^2
    +\frac {1}{2} ( \mid {\cal M}_{u\bar {u}\bar {d}
                               \to u\bar {u}\bar {d}} \mid^2
+ \mid {\cal M}_{u\bar {d}\bar {u} \to u\bar {d}\bar {u}} \mid^2 )
+ \frac {1}{4}
  \mid {\cal M}_{u\bar {d}\bar {d} \to u\bar {d}\bar {d}} \mid^2 ]
         \nonumber    \\
\fl
~~~ \times \left. [f_{q1}f_{q2}f_{q3}(1-f_{q4})(1-f_{q5})(1-f_{q6})
-f_{q4}f_{q5}f_{q6}(1-f_{q1})(1-f_{q2})(1-f_{q3})] \right\} ,
         \nonumber    \\
\end{eqnarray}
where $\rm \vec {v}_1$ is the velocity of the massless gluon or up quark;
the colour-spin degeneracy factors are $g_G=16$ for the gluon and $g_Q=6$
for the quark; $p_1$ and $p_2$ ($p_1$, $p_2$ and $p_3$) denote the four-momenta
of the two (three) initial particles, and $p_3$ and $p_4$ ($p_4$, $p_5$ and 
$p_6$) of the two (three) final particles in 2-to-2 (3-to-3) scattering; and
$E_i$ is the energy component of $p_i$. The squared amplitudes of order 
$\alpha^2_{\rm s}$ for the 
elastic 2-to-2 scattering, $\mid {\cal M}_{gg \to gg} \mid^2$, 
$\mid {\cal M}_{gu \to gu} \mid^2$, etc., can be found in Refs. \cite{cs,ckr}.
The squared amplitude of order $\alpha_{\rm s}^4$ for the elastic
$ggg$ scattering, $\mid {\cal M}_{ggg \to ggg} \mid^2$, 
was obtained in the work of Ref. \cite{xmxu3}.
Since the elastic gluon-gluon-quark scattering involves a lot more Feynman
diagrams than the elastic $gqq$ scattering and the elastic
$gq\bar q$ scattering, it will take two years to derive the squared
amplitude for the elastic gluon-gluon-quark scattering. For the time being, we
have to give up the elastic scattering of both $ggq$ and $gg\bar q$, i.e. set
\begin{displaymath}
{\cal M}_{ugg \to ugg}={\cal M}_{ggu \to ggu}={\cal M}_{ggd \to ggd}
={\cal M}_{gg\bar u \to gg\bar u}={\cal M}_{gg\bar d \to gg\bar d}=0.
\end{displaymath}
The squared amplitude for the elastic gluon-antiquark-antiquark scattering
equals the one for the elastic $gqq$ scattering, for example,
\begin{displaymath}
\mid {\cal M}_{g\bar {u}\bar {u} \to g\bar {u}\bar {u}} \mid^2 
= \mid {\cal M}_{guu \to guu} \mid^2,~~~~~
\mid {\cal M}_{g\bar {u}\bar {d} \to g\bar {u}\bar {d}} \mid^2 
= \mid {\cal M}_{gud \to gud} \mid^2.
\end{displaymath}
The squared amplitude $\mid {\cal M}_{qqg \to qqg} \mid^2$ is obtained from 
$\mid {\cal M}_{gqq \to gqq} \mid^2$ by the replacement of 
$p_1 \leftrightarrow p_3$
and $\mid {\cal M}_{q\bar {q}g \to q\bar {q}g} \mid^2$ from 
$\mid {\cal M}_{gq\bar q \to gq\bar q} \mid^2$ by the replacement of 
$p_1 \to p_3$, 
$p_2 \to p_1$ and $p_3 \to p_2$. The squared amplitudes for the elastic 
quark-quark-quark or antiquark-antiquark-antiquark scattering and for the
elastic quark-quark-antiquark or quark-antiquark-antiquark scattering were
obtained in the work of Ref. \cite{xmxu1} and of Ref. \cite{xmxu2}, 
respectively. Similar equations for down quarks, up antiquarks and down 
antiquarks can be established.

The squared amplitude, $\mid {\cal M}_{gqq \to gqq} \mid^2$ for
$g(p_1)+q(p_2)+q(p_3) \to g(p_4)+q(p_5)+q(p_6)$ or $\mid 
{\cal M}_{gq\bar {q} \to gq\bar {q}} \mid^2$ for
$g(p_1)+q(p_2)+\bar {q}(-p_3) \to g(p_4)+q(p_5)+\bar {q}(-p_6)$, is the sum of
the individually squared amplitudes of the diagrams in the three classes and 
interference terms of different diagrams. The sum of the two interference
terms of a diagram in the first or third class and a diagram in
the second class equals zero. Therefore, there is no interference between the
first or third class and the second class. Any
interference term between a diagram in the first class and a diagram in the
third class has a very long expression and is thus not shown here.
Examples of the individually squared amplitudes are the spin- and color-summed 
squared amplitudes for the diagrams ${\rm D}_{\sim {\rm LM}}$ and 
${\rm E}_{\sim {\rm UU}}$ which, as shown below,  
include the average over the spin and color states of the initial particles,
\begin{eqnarray}
\frac {1}{8} \frac {1}{72} 
& &
\sum\limits_{\rm {spins,colors}}
\mid {\cal M}_{{\rm D}_{\sim {\rm LM}}} \mid^2
= \frac {32 {\rm g}_{\rm s}^8}{9}\frac {1}{18}
(  s_{12}u_{16}u_{35}^2  -s_{12}u_{16}u_{34}^2  -s_{12}u_{16}u_{26}u_{35} 
                   \nonumber   \\
& &
  +s_{12}u_{16}u_{26}u_{34}  -s_{12}u_{16}^2u_{35}  +s_{12}u_{16}^2u_{34}
  -s_{12}s_{31}u_{35}^2
                   \nonumber   \\
& &
  +s_{12}s_{31}u_{34}^2  +2s_{12}s_{31}u_{26}u_{35} -s_{12}s_{31}u_{26}^2
  +2s_{12}s_{31}u_{24}u_{35}
                   \nonumber   \\
& &
  +2s_{12}s_{31}u_{24}u_{34}  -2s_{12}s_{31}u_{24}u_{26}      
  -2s_{12}s_{31}u_{16}u_{34}  -2s_{12}s_{31}u_{16}u_{24}
                   \nonumber   \\
& &
  +s_{12}s_{31}u_{16}^2   -2s_{12}s_{31}u_{15}u_{35}  
  -2s_{12}s_{31}u_{15}u_{34} +2s_{12}s_{31}u_{15}u_{26}
                   \nonumber   \\
& &
  +2s_{12}s_{31}u_{15}u_{16}  -s_{12}s_{31}^2u_{35}
  -s_{12}s_{31}^2u_{34}  +s_{12}s_{31}^2u_{26}     
                   \nonumber   \\
& &
  +s_{12}s_{31}^2u_{16}  +s_{12}s_{23}s_{31}u_{35} 
  +s_{12}s_{23}s_{31}u_{34}  -s_{12}s_{23}s_{31}u_{26}
                   \nonumber   \\
& &
  -s_{12}s_{23}s_{31}u_{16}  -s_{12}^2u_{16}u_{35}  +s_{12}^2u_{16}u_{34}
  +s_{12}^2s_{31}u_{35} 
                   \nonumber   \\
& &
  -s_{12}^2s_{31}u_{34}  -s_{12}^2s_{31}u_{26}
  -2s_{12}^2s_{31}u_{24} +s_{12}^2s_{31}u_{16}
                   \nonumber   \\
& &
  +2s_{12}^2s_{31}u_{15} +s_{12}^2s_{31}^2 -s_{12}^2s_{23}s_{31} )
                   \nonumber   \\
& &
/[s_{12}(s_{12}+u_{16}+u_{26}-u_{34}-u_{35})(-s_{31}-s_{23}-u_{34}-u_{35})]^2,
\end{eqnarray}
\begin{eqnarray}
\frac {1}{8} \frac {1}{72} 
& &
\sum\limits_{\rm {spins,colors}}
\mid {\cal M}_{{\rm E}_{\sim {\rm UU}}} \mid^2
= \frac {32 {\rm g}_{\rm s}^8}{9}\frac {1}{18}
(  -s_{31}u_{15}u_{26}u_{35} +s_{31}u_{15}u_{26}^2  
                   \nonumber   \\
& &
 +s_{31}u_{15}u_{24}u_{35}  -s_{31}u_{15}u_{24}^2  -s_{31}u_{15}^2u_{26}  
 +s_{31}u_{15}^2u_{24}  -s_{31}^2u_{15}u_{26}
                   \nonumber   \\
& &
  +s_{31}^2u_{15}u_{24}  -2s_{23}u_{15}u_{16}u_{35}
  +2s_{23}u_{15}u_{16}u_{26}  +2s_{23}u_{15}u_{16}u_{24}
                   \nonumber   \\
& &
  -s_{23}u_{15}^2u_{35}  +s_{23}u_{15}^2u_{26} +s_{23}u_{15}^2u_{24}
  -2s_{23}u_{15}^2u_{16}
                   \nonumber   \\
& &
  -s_{23}u_{15}^3   -s_{23}s_{31}u_{15}u_{35}  +s_{23}s_{31}u_{15}u_{26}
  +s_{23}s_{31}u_{15}u_{24}
                   \nonumber   \\
& &
  -2s_{23}s_{31}u_{15}u_{16}  -2s_{23}s_{31}u_{15}^2
  -s_{23}s_{31}^2u_{15}  -s_{12}s_{23}u_{15}u_{35}
                   \nonumber   \\
& &
  +s_{12}s_{23}u_{15}u_{26}  +s_{12}s_{23}u_{15}u_{24}  -s_{12}s_{23}u_{15}^2
  -s_{12}s_{23}s_{31}u_{15} )
                   \nonumber   \\
& &
/[u_{15}(s_{31}+u_{15}-u_{24}-u_{26}+u_{35})(-s_{23}-u_{24}-u_{26}+u_{15})]^2,
\end{eqnarray}
where ${\rm g}_{\rm s}$ is the gauge coupling constant and
nine variables are defined as
$s_{12}=(p_1+p_2)^2$, $s_{23}=(p_2+p_3)^2$, $s_{31}=(p_3+p_1)^2$,
$u_{15}=(p_1-p_5)^2$, $u_{16}=(p_1-p_6)^2$,
$u_{24}=(p_2-p_4)^2$, $u_{26}=(p_2-p_6)^2$,
$u_{34}=(p_3-p_4)^2$ and $u_{35}=(p_3-p_5)^2$. Squared amplitudes for 2-to-4 
processes with two gluons and four quarks were obtained from helicity 
amplitudes in Ref. \cite{gk} and can also be derived from Fortran code named 
CompHEP \cite{CompHEP}. One recent review on on-shell methods of scattering
amplitudes in perturbative QCD has been given in Ref. \cite {bdk}.
Momenta of five partons among the six partons are randomly generated and the
momentum of the other parton is given by energy-momentum conservation. With
such a set of six momenta 
numerical values of the expressions in Eqs. (4) and (5) agree
with numerical results of the 2-to-4 processes presented in Refs. 
\cite{gk,CompHEP} by reversing the momentum of one final parton.
Such agreement is also established for any other set of momenta randomly 
generated.

Hadronic matter and quark-gluon plasma exist below and above the critical 
temperature of the QCD phase transition, respectively. 
Due to the medium screening, the gauge 
coupling constant in quark-gluon plasma is smaller than in hadronic matter and 
at temperatures high enough
is so small that perturbative QCD can be applied. The gauge
coupling constant decreases while the temperature increases \cite {kapusta,ls}.
The higher the temperature is, the better any perturbative expansion converges.
While the temperature is near the critical temperature, the perturbative 
expansion breaks down; but no one has determined the breakdown temperature 
regime.

\section{Numerical solutions and discussions}

Time dependence of distributions of gluons, quarks and antiquarks is determined
by the transport equations which have the initial condition at $t=0.2$ fm/$c$
generated by HIJING \cite{wwwgg} for central Au-Au collisions at 
$\sqrt {s_{NN}}= 200$ GeV
and expressed in the form \cite{lmw}
\begin{equation}
f(k_\bot,y,r,z,t)=\frac {1}{16\pi R_A^2}g(k_\bot,y)\frac
{e^{-(z-t\tanh y)^2/2\Delta_k^2}}{\sqrt {2\pi}\Delta_k},
\end{equation}
with
\begin{displaymath}
\Delta_k \approx \frac {2}{k_\bot \cosh y},
\end{displaymath}
and
\begin{displaymath}
g(k_\bot,y)=\frac {(2\pi)^3}{k_\bot \cosh y} \frac {dN}{dyd^2k_\bot},
\end{displaymath}
where $R_A$, $k_\bot$, $y$, $t$, $z$ and $r$ are the gold nucleus radius, 
transverse momentum, rapidity, time, coordinate in the longitudinal
direction and radius in the transverse direction, respectively. The gluon and 
the quark have different $dN/dyd^2k_\bot$. One thousand and five
hundred gluons within $-0.3<z<0.3$ fm and $r<R_A$ are created from the 
distribution by the rejection method. 
Two hundred and fifty quarks or antiquarks of the
up or down flavor are created in the same region.

Scattering of two partons happens when the two partons have the closest
distance less than the square root of the ratio of cross section for 2-to-2
scattering to $\pi$. The cross section for $gg \to gg$ is
\begin{equation}
\fl
\sigma_{gg \to gg}
= \frac {{\rm g}_{\rm s}^4}{16\pi s^2} \frac {9}{4} \left[ 
\frac {17s^3+66\mu_D^2s(s+\mu_D^2)}{6(s+2\mu_D^2)^2}
+ \frac {2s(s+2\mu_D^2)^2}{\mu_D^2(s+\mu_D^2)}
+ 2(s+2\mu_D^2) \ln \frac {\mu_D^2}{s+\mu_D^2} \right] ,
\end{equation}
and the cross section for $gq \to gq$ or $g\bar {q} \to g\bar q$ is
\begin{equation}
\fl
\sigma_{gq \to gq}=\sigma_{g\bar {q} \to g\bar {q}}
= \frac {{\rm g}_{\rm s}^4}{16\pi s^2} \left[ \frac {11}{9}s
+ \frac {14}{9}(s+2\mu_D^2)\ln \frac {\mu_D^2}{s+\mu_D^2} 
+ \frac {2s(s+2\mu_D^2)^2}{\mu_D^2(s+\mu_D^2)} \right] ,
\end{equation}
where $s$ is the square of the total energy of two colliding particles in the
center-of-momentum system;
$\mu_D$ is the screening mass formulated in Refs. \cite{bmw,bms,kk} and 
is used to regularize propagators. 
The coupling constant  $\alpha_{\rm s}={\rm g}_{\rm s}^2/4\pi=0.5$ 
is taken in finding solutions of the transport equations. The cross
section for the elastic $qq$, $q\bar q$ or $\bar {q}\bar {q}$ scattering can be
found in Ref. \cite{xmxu2} where the fraction 8/9 in Eq. (5) should be replaced
by 4/9.

Scattering of three partons occurs if the three partons
are in a sphere of which the center is at the center-of-mass of the three 
partons and of which the radius $r_{\rm hs}$ is \cite{xmxu2}
\begin{eqnarray}
\pi r_{\rm hs}^2
& = & \frac {1}{m} \int \frac {d^3p_4}{(2\pi)^32E_4}
\frac {d^3p_5}{(2\pi)^32E_5} \frac {d^3p_6}{(2\pi)^32E_6}
         \nonumber    \\
& &
\times (2\pi)^4 \delta^4(p_1+p_2+p_3-p_4-p_5-p_6)
\mid {\cal M}_{3 \to 3} \mid^2,
\end{eqnarray}
where $m=1$ for $gud \to gud$, $g\bar {u}\bar {d} \to 
g\bar {u}\bar {d}$, $gu\bar {u} \to gu\bar {u}$, $gd\bar {d} \to gd\bar {d}$, 
$gu\bar {d} \to gu\bar {d}$ or $gd\bar {u} \to gd\bar {u}$, and $m=4$ for 
$guu \to guu$, $gdd \to gdd$, $g\bar {u}\bar {u} \to g\bar {u}\bar {u}$ or 
$g\bar {d}\bar {d} \to g\bar {d}\bar {d}$. 

As shown by Eq. (6), the momentum distribution in the longitudinal direction is
very different from the one in the transverse direction, but
gluons, quarks and antiquarks at the initial time have similar anisotropy.
Gluon distribution functions in all directions overlap at the time 
$t=0.68~{\rm fm}/c$ which corresponds to a thermalization time of the order of
0.48 fm/$c$ and quark distribution functions at $t=1.56~{\rm fm}/c$ which gives
a thermalization time of about 1.36 fm/$c$. These distribution
functions are plotted in Figs. 15 and 16, respectively. Solid curves in Figs. 
15 and 16 stand for the J$\rm \ddot u$ttner distribution
\begin{equation}
f_g(\vec {p})=\frac {\lambda_g}{{\rm e}^{\mid \vec {p}\mid/T}-\lambda_g},
\end{equation}
with $T=0.5$ GeV and $\lambda_g=0.3$ for gluon matter at $t=0.68~{\rm fm}/c$,
and
\begin{equation}
f_q(\vec {p})=\frac {\lambda_q}{{\rm e}^{\mid \vec {p}\mid/T}+\lambda_q},
\end{equation}
with $T=0.3$ GeV and $\lambda_q=0.3$ for quark matter at $t=1.56~{\rm fm}/c$,
respectively.

The solutions of the transport equations indicate that
gluon matter thermalizes rapidly and quark matter thermalizes slowly. How fast
thermalization is depends on squared amplitudes and distribution functions
\cite{xmxu4}.
Calculations in perturbative QCD show that 
$\mid {\cal M}_{gu \to gu} \mid^2=\mid {\cal M}_{gd \to gd} \mid^2=
\mid {\cal M}_{g\bar {u} \to g\bar {u}} \mid^2
=\mid {\cal M}_{g\bar {d} \to g\bar {d}} \mid^2$, then the term
$g_Q ( \mid {\cal M}_{gu \to gu} \mid^2
+ \mid {\cal M}_{gd \to gd} \mid^2
+ \mid {\cal M}_{g\bar {u} \to g\bar {u}} \mid^2
+ \mid {\cal M}_{g\bar {d} \to g\bar {d}} \mid^2 )$ in Eq. (2) equals 
$24\mid {\cal M}_{gu \to gu} \mid^2$ which is near
$g_G \mid {\cal M}_{ug \to ug} \mid^2$ in Eq. (3). Therefore, variation of the
gluon distribution function caused by elastic scattering of both $gq$ and
$g\bar q$ is near variation of the 
quark distribution function caused by the same
scattering. Numerical calculations lead to a similar conclusion about variation
of the gluon and quark distribution functions caused by elastic scattering of
$gqq$, $gq\bar q$ and $g\bar {q}\bar q$. Therefore, the 
difference between the change of gluon distribution and the change of quark
distribution or the difference between the 
thermalization time of gluon matter and the one of quark matter
is mainly given by the elastic $ggg$ scattering and the elastic 
$qqq$ scattering. Then we need to see the point of view from the 
$ggg$ and $qqq$ scattering.
This is accomplished by the four aspects: (1) the gluon distribution function 
$f_{gi}$ is about 2 times the quark
distribution function $f_{qi}$; (2) the maximum of 
$\mid {\cal M}_{gg \to gg} \mid^2 f_{gi}f_{gj}$ in Eq. (2) is an order of
magnitude larger than that of
$(\frac {1}{2} \mid {\cal M}_{uu \to uu} \mid^2
+ \mid {\cal M}_{ud \to ud} \mid^2 
+ \mid {\cal M}_{u\bar {u} \to u\bar {u}} \mid^2
+ \mid {\cal M}_{u\bar {d} \to u\bar {d}} \mid^2 ) f_{qi}f_{qj}$ in Eq. (3);
(3) the maximum of $\mid {\cal M}_{ggg \to ggg} \mid^2$
is two orders of magnitude larger than that of 
$\mid {\cal M}_{qqq \to qqq} \mid^2$, 
$\mid {\cal M}_{qq\bar {q} \to qq\bar {q}} \mid^2$ or
$\mid {\cal M}_{q\bar {q}\bar {q} \to q\bar {q}\bar {q}} \mid^2$;
(4) the factor $g_{G}^2f_{gi}f_{gj}f_{gk}/12$ is over four times the factor
$g_{Q}^2f_{qi}f_{qj}f_{qk}$. Finally, we understand that the 
rapid thermalization 
of gluon matter and the slow thermalization of quark matter result mainly from
the fact that the elastic $gg$ ($ggg$) scattering has a larger
squared amplitude than the elastic $qq$ or $q\bar q$ ($qqq$ or $qq\bar q$) 
scattering and gluon matter is denser than quark matter. 

The squared amplitude for the elastic $gq$ ($gqq$ or $gq\bar q$) 
scattering is comparable to the one for the 
elastic $qq$ or $q\bar q$ ($qqq$ or $qq\bar q$) scattering.
To get a clear understanding of contributions of new terms of the
elastic scattering of $gqq$ and $gq\bar q$ in Eqs. (2) and (3),
we approximate the factors, $1+f_{gi}$ and $1-f_{qi}$, by 1.
In Eq. (2) the maximum of the new term
$
g_Q ( \mid {\cal M}_{gu \to gu} \mid^2
+ \mid {\cal M}_{gd \to gd} \mid^2
+ \mid {\cal M}_{g\bar {u} \to g\bar {u}} \mid^2
+ \mid {\cal M}_{g\bar {d} \to g\bar {d}} \mid^2 ) (f_{g1}f_{q2}-f_{g3}f_{q4})
$
is about half of the maximum of the term
$
\frac {g_G}{2} \mid {\cal M}_{gg \to gg} \mid^2 (f_{g1}f_{g2}-f_{g3}f_{g4}),
$ and the maximum of the new term
$ 
g_Q^2 [\frac {1}{4} \mid {\cal M}_{guu \to guu} \mid^2
+\frac {1}{2} ( \mid {\cal M}_{gud \to gud} \mid^2
              + \mid {\cal M}_{gdu \to gdu} \mid^2 )
+\frac {1}{4} \mid {\cal M}_{gdd \to gdd} \mid^2
    + \mid {\cal M}_{gu\bar {u} \to gu\bar {u}} \mid^2
    + \mid {\cal M}_{gu\bar {d} \to gu\bar {d}} \mid^2
    + \mid {\cal M}_{gd\bar {u} \to gd\bar {u}} \mid^2
    + \mid {\cal M}_{gd\bar {d} \to gd\bar {d}} \mid^2
    +\frac {1}{4} \mid {\cal M}_{g\bar {u}\bar {u}
                             \to g\bar {u}\bar {u}} \mid^2
    +\frac {1}{2} ( \mid {\cal M}_{g\bar {u}\bar {d}
                               \to g\bar {u}\bar {d}} \mid^2 
    + \mid {\cal M}_{g\bar {d}\bar {u} \to g\bar {d}\bar {u}} \mid^2 )
+ \frac {1}{4} \mid {\cal M}_{g\bar {d}\bar {d} \to g\bar {d}\bar {d}} \mid^2 ]
(f_{g1}f_{q2}f_{q3} -f_{g4}f_{q5}f_{q6})
$ 
is about one-fifth of the maximum of     
$
\frac {g_G^2}{12} \mid {\cal M}_{ggg \to ggg} \mid^2
(f_{g1}f_{g2}f_{g3}-f_{g4}f_{g5}f_{g6}).
$
Therefore, the new terms provide small contributions to thermalization of 
gluon matter. In Eq. (3) the maximum of the new term
$
g_G \mid {\cal M}_{ug \to ug} \mid^2 (f_{q1}f_{g2}-f_{q3}f_{g4}) 
$ is about 2 times the maximum of the term
$
g_Q (\frac {1}{2} \mid {\cal M}_{uu \to uu} \mid^2
+ \mid {\cal M}_{ud \to ud} \mid^2 
+ \mid {\cal M}_{u\bar {u} \to u\bar {u}} \mid^2
+ \mid {\cal M}_{u\bar {d} \to u\bar {d}} \mid^2 )
(f_{q1}f_{q2}-f_{q3}f_{q4}),
$ and the maximum of the new term
$
g_Qg_G ( \frac {1}{2} \mid {\cal M}_{uug \to uug} \mid^2
+\mid {\cal M}_{udg \to udg} \mid^2
+\mid {\cal M}_{u\bar {u}g \to u\bar {u}g} \mid^2
+\mid {\cal M}_{u\bar {d}g \to u\bar {d}g} \mid^2 )
(f_{q1}f_{q2}f_{g3}-f_{q4}f_{q5}f_{g6})
$ is near the maximum of
$
g_Q^2 [\frac {1}{12} \mid {\cal M}_{uuu \to uuu} \mid^2
+\frac {1}{4} ( \mid {\cal M}_{uud \to uud} \mid^2
              + \mid {\cal M}_{udu \to udu} \mid^2 )
+\frac {1}{4} \mid {\cal M}_{udd \to udd} \mid^2
    +\frac {1}{2} \mid {\cal M}_{uu\bar {u} \to uu\bar {u}} \mid^2
    +\frac {1}{2} \mid {\cal M}_{uu\bar {d} \to uu\bar {d}} \mid^2
            + \mid {\cal M}_{ud\bar {u} \to ud\bar {u}} \mid^2
            + \mid {\cal M}_{ud\bar {d} \to ud\bar {d}} \mid^2
    +\frac {1}{4} \mid {\cal M}_{u\bar {u}\bar {u}
                             \to u\bar {u}\bar {u}} \mid^2
    +\frac {1}{2} ( \mid {\cal M}_{u\bar {u}\bar {d}
                               \to u\bar {u}\bar {d}} \mid^2
+ \mid {\cal M}_{u\bar {d}\bar {u} \to u\bar {d}\bar {u}} \mid^2 )
+ \frac {1}{4}
  \mid {\cal M}_{u\bar {d}\bar {d} \to u\bar {d}\bar {d}} \mid^2 ]
  (f_{q1}f_{q2}f_{q3}-f_{q4}f_{q5}f_{q6}).
$ Hence, the new terms have comparable contributions to thermalization of 
quark matter. Governed by the elastic scattering of $qq$, $q\bar q$, 
$qqq$, $qq\bar q$ and $q\bar {q}\bar {q}$, a 
thermalization time of the order of 1.55 fm/$c$ was obtained in
Ref. \cite{xmxu2} for quark matter with the same initial distribution as
Eq. (6). Hence, the elastic scattering of  $gq$, $gqq$ and $gq\bar q$ 
shortens the thermalization time of quark matter by the 
amount 0.19 fm/$c$. About half of the amount is a consequence of the elastic 
$gqq$ and $gq\bar q$ scattering. The elastic $gq$ scattering, the elastic $gqq$
scattering and the elastic $gq\bar q$ scattering have different contributions 
in shortening the thermalization time. Taking up-quark matter as an example,
amounts by which the thermalization time is shortened by relevant elastic 
scattering are listed in Table 1. The amount by which the thermalization time
is shortened by the elastic $gq\bar q$ scattering is larger than by $gqq$.

In Ref. \cite{xmxu3} one thousand gluons are generated from a distribution that
is homogeneous in coordinate space but anisotropic in momentum space. Gluon
matter is controlled to evolve in the longitudinal direction and is governed by
the elastic $ggg$ scattering. The resultant fugacity and temperature are 0.065 
and 
0.75 GeV, respectively. In Ref. \cite{xmxu1} six hundred and sixty-six quarks 
are generated from a distribution similar to that in Ref. \cite{xmxu3}. Quark
matter is also controlled to evolve in the longitudinal direction and is 
governed by the elastic $qqq$ scattering. The resultant fugacity and 
temperature 
are 0.04 and 0.59 GeV, respectively. Due to the restriction of longitudinal 
expansion, the two fugacities are small and the two temperatures are high.
In Ref. \cite{xmxu2} five hundred quarks and five hundred antiquarks 
are created from the same distribution as that in Eq. (6). 
Governed by the elastic scattering of $qqq$, $qq\bar q$, $q\bar {q}\bar q$ and 
$\bar {q}\bar {q}\bar q$, quark matter and antiquark matter evolve in both the 
transverse direction and the longitudinal direction.
The resultant fugacity and temperature are 0.31 and 0.27 GeV, respectively. 
Without the restriction of longitudinal expansion, the fugacities in Ref.
\cite{xmxu2} and in the present work are not small and the temperatures are not
higher than those in Refs. \cite{xmxu3} and \cite{xmxu1}. The fugacity and 
temperature obtained in the present work must be different from those in Ref.
\cite{xmxu2} since the present work involves gluon matter that is absent in 
Ref. \cite{xmxu2}.

Even though the study and application of elastic 4-to-4 scattering is not the
purpose of the present work, we still can know the occurrence probability of
the 4-parton scattering and it has been shown in Ref. \cite {xmxu2}.
One thousand and five hundred partons were generated from the same
distribution as that in Eq. (6) within $-0.3~{\rm fm} < z < 0.3~{\rm fm}$ and 
$r<R_A$. Given an interaction range of 0.62 fm, the 2-parton scattering has the
occurrence probability of 30\%, the 3-parton scattering 20\%, and
the 4-parton scattering 14.6\%. Therefore, the elastic 4-parton scattering
is expected to give a smaller contribution to thermalization than the
elastic 3-parton scattering. Interaction of partons in a sphere with a radius 
the same as the interaction range takes place and the number of partons in 
such a sphere is at
most 14. This means that at most 14-parton scattering is allowed. As a 
consequence, the occurrence probability of 15-parton scattering is zero and the
one of the 14-parton scattering is very small. The occurrence probabilities for
5-parton, 6-parton, 7-parton, 8-parton, 9-parton, 10-parton and 11-parton
scattering are 11\%, 9\%, 7.5\%, 4.4\%, 2.3\%, 0.9\% and 0.2\%, respectively. 
The occurrence probabilities for 12-parton scattering and 13-parton 
scattering are negligible. The sum of the 
occurrence probabilities from the 2-parton scattering through the 14-parton 
scattering equals 1 and the occurrence probabilities form a convergent series.

\section{Summary}
We have established the transport equations that include the squared amplitudes
for the elastic scattering of $gqq$, $gq\bar q$ and $g\bar {q}\bar q$. 
The squared amplitudes are derived at the tree level of the scattering in
perturbative QCD and are expressed in terms of the nine Lorentz-invariant 
momentum variables. The elastic scattering of $gqq$, $gq\bar q$ and $g\bar {q}
\bar {q}$ shortens the thermalization time of quark matter as well as antiquark
matter. This is an effect of the three-body scattering while the number density
is high. In not only quark-gluon
matter with a high number density but also a very dense scalar field system,
the elastic 3-to-3 scattering plays a significant role \cite{cm}.

\ack
This work was supported in part by the National Natural Science Foundation of
China under Grant No. 10675079 and in part by Shanghai Leading Academic 
Discipline Project (project number S30105).

\newpage

\begin{table}[htbp]
\centering \caption{Amounts by which the thermalization time is shortened.}
\label{massvalue}
\begin{tabular*}{16.5cm}{@{\extracolsep{\fill}}cccc}
  \hline
  $gu \to gu$  & $guu \to guu, gud \to gud$ & 
  $gu\bar {u} \to gu\bar {u}, gu\bar {d} \to gu\bar {d}$ \\
  \hline
  0.091 fm/$c$ & 0.046 fm/$c$ & 0.053 fm/$c$ \\
  \hline
\end{tabular*}
\end{table}

\newpage
\begin{figure}
  \centering
    \includegraphics[width=42mm,height=65mm,angle=0]{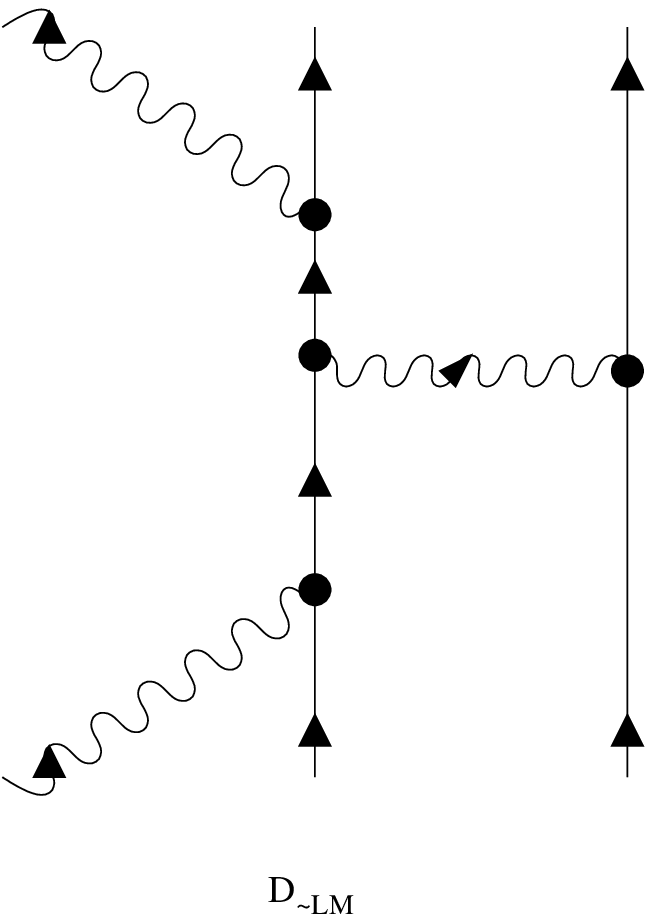}
      \hspace{1.2cm}
    \includegraphics[width=42mm,height=65mm,angle=0]{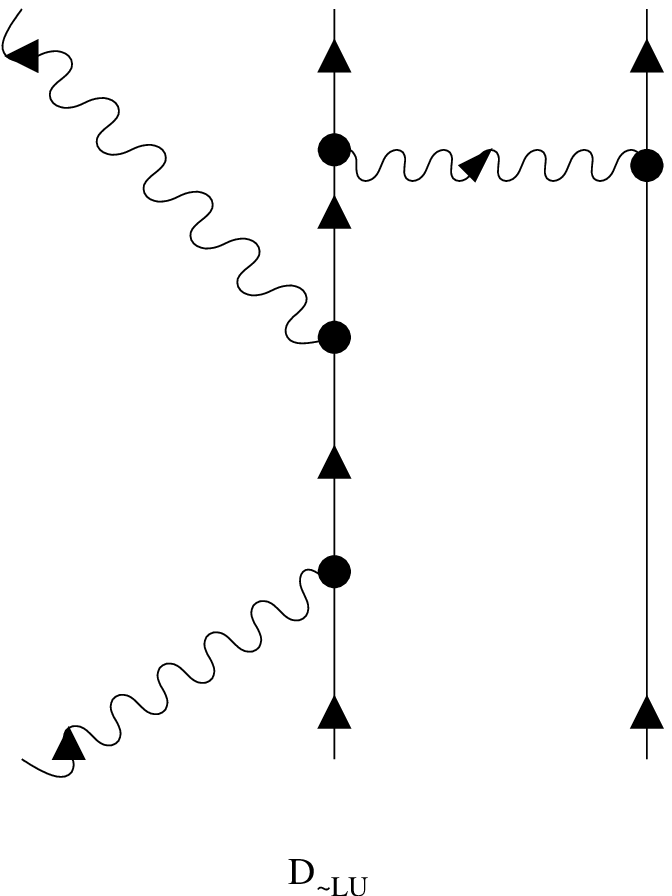}
      \hspace{1.2cm}
    \includegraphics[width=42mm,height=65mm,angle=0]{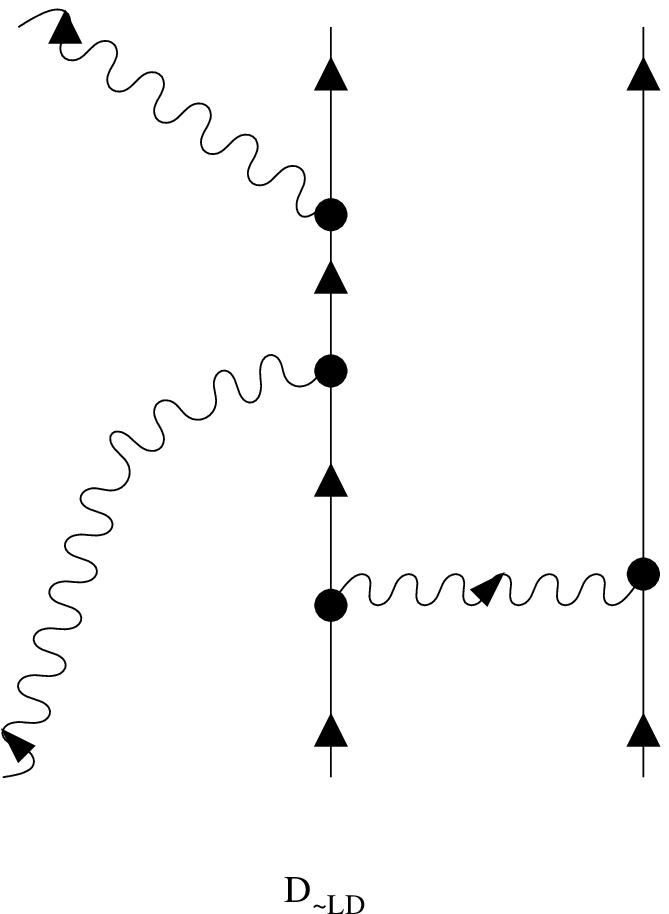}
      \vskip 26pt
    \includegraphics[width=42mm,height=65mm,angle=0]{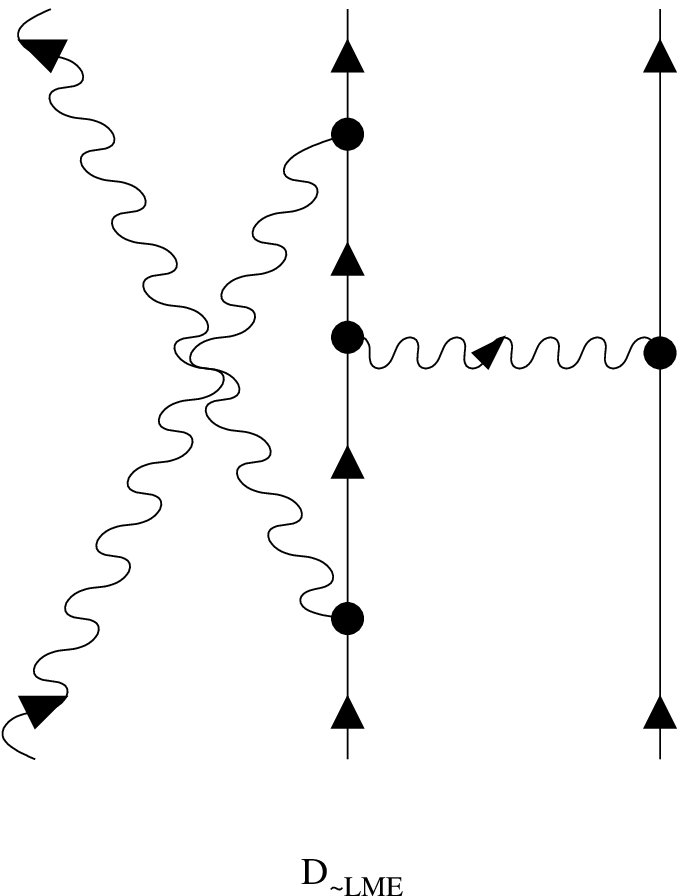}
      \hspace{1.2cm}
    \includegraphics[width=42mm,height=65mm,angle=0]{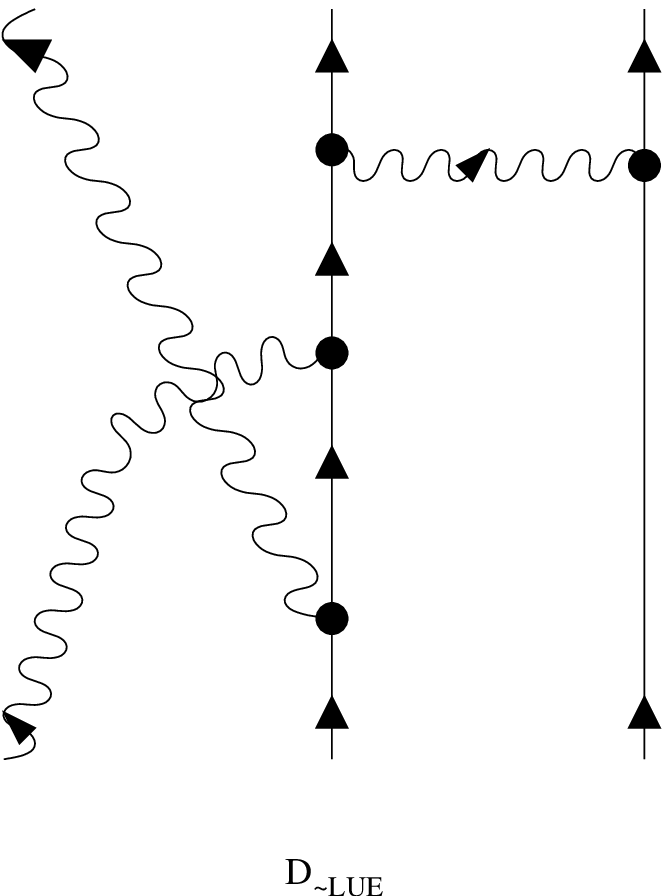}
      \hspace{1.2cm}
    \includegraphics[width=42mm,height=65mm,angle=0]{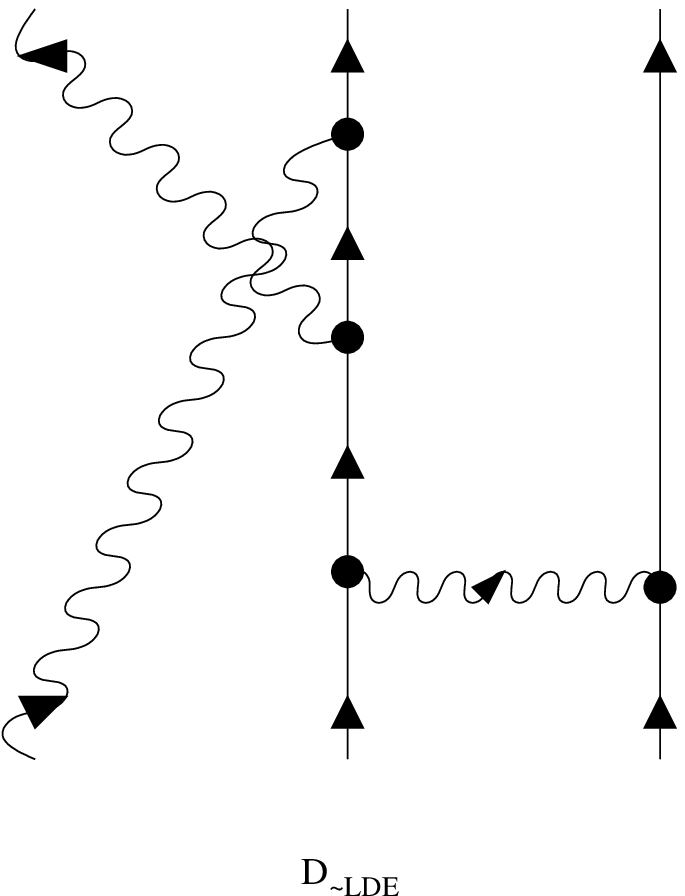}
\caption{Elastic gluon-quark-quark scattering.}
\label{fig1}
\end{figure}

\newpage
\begin{figure}
  \centering
    \includegraphics[width=42mm,height=65mm,angle=0]{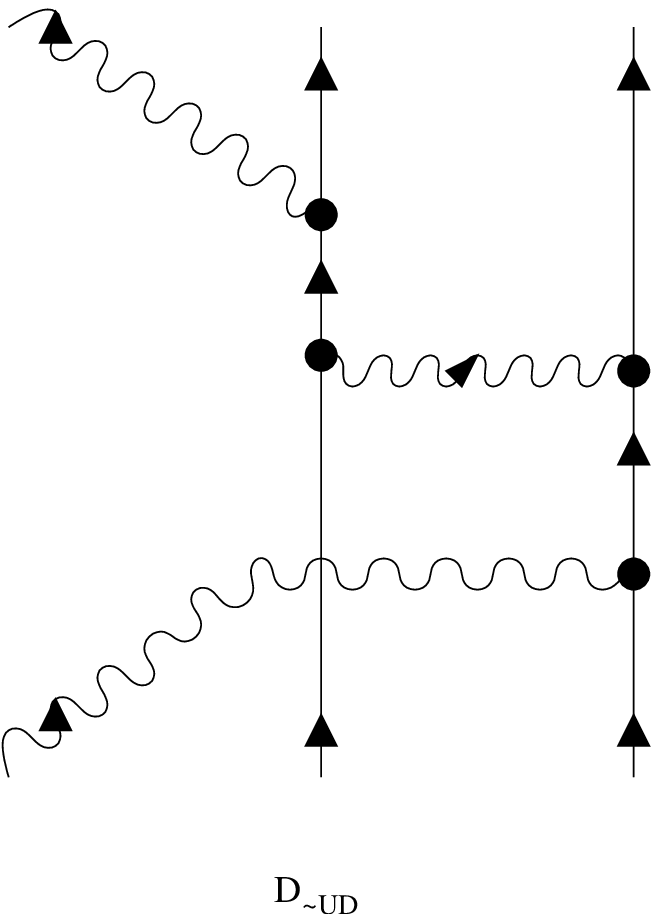}
      \hspace{1.2cm}
    \includegraphics[width=42mm,height=65mm,angle=0]{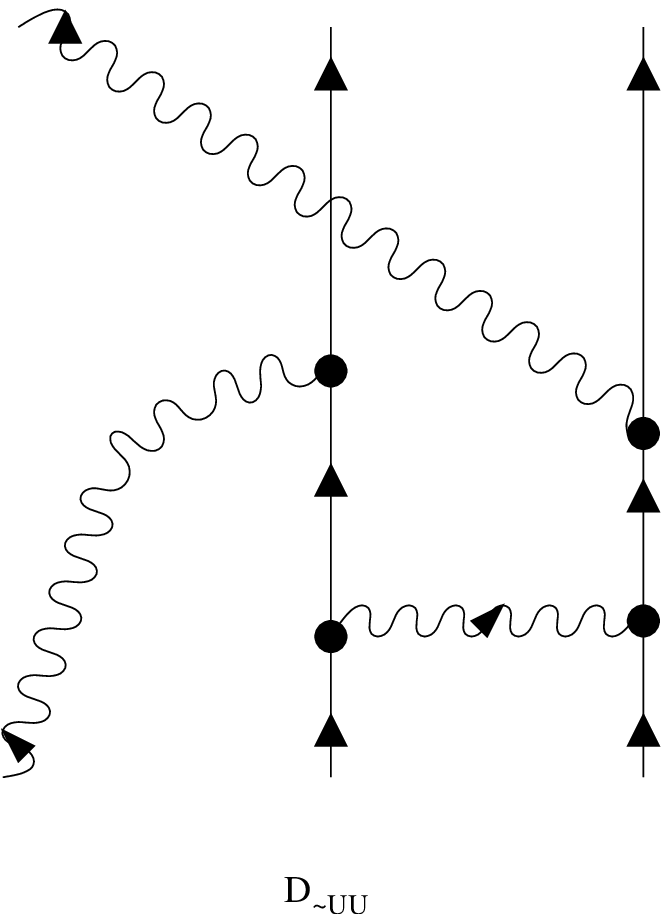}
      \vskip 26pt
    \includegraphics[width=42mm,height=65mm,angle=0]{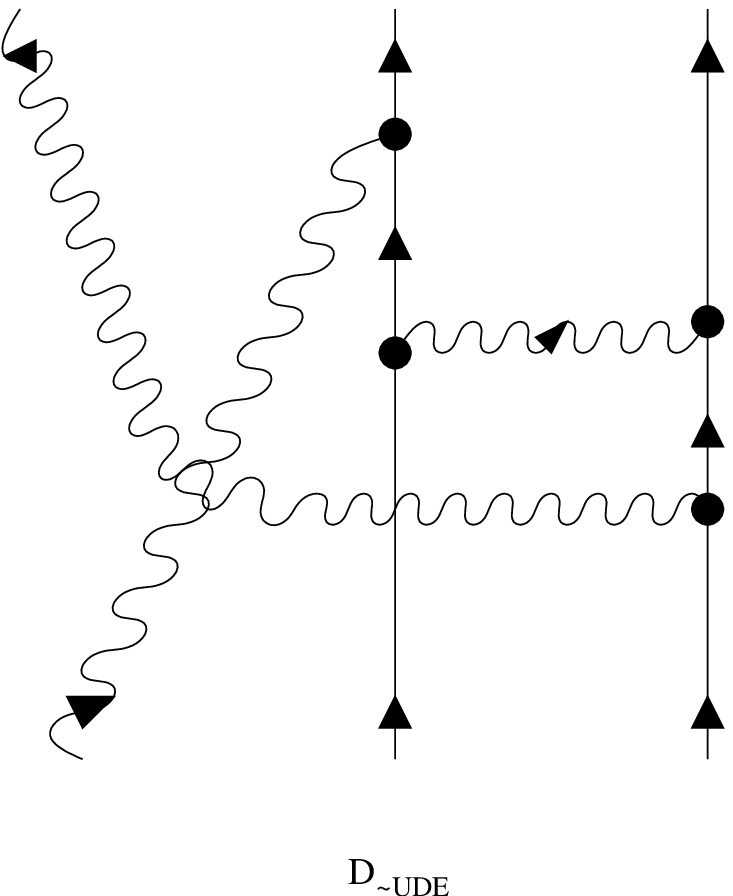}
      \hspace{1.2cm}
    \includegraphics[width=42mm,height=65mm,angle=0]{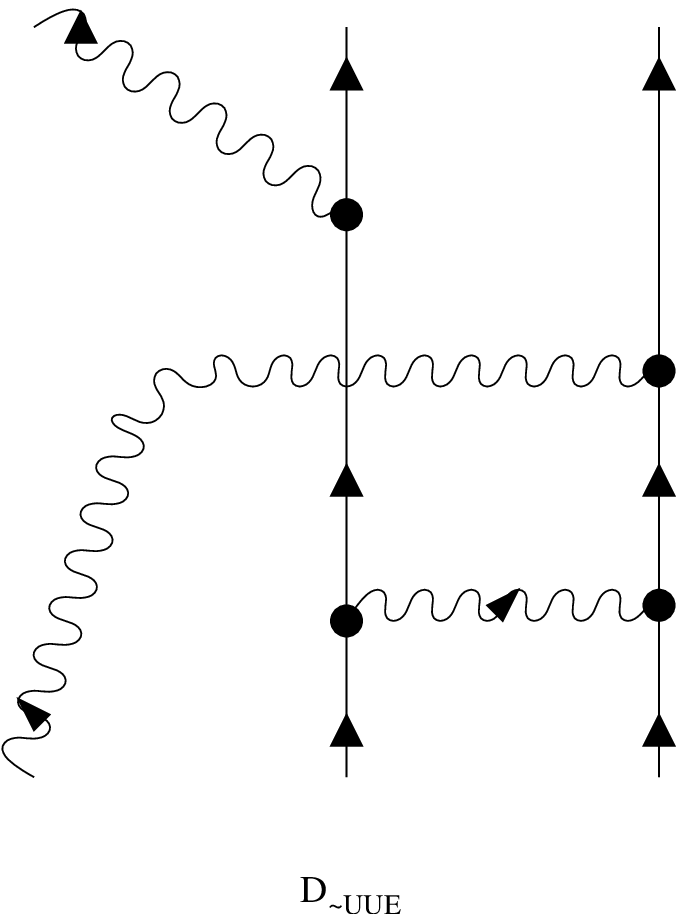}
\caption{Elastic gluon-quark-quark scattering.}
\label{fig2}
\end{figure}

\newpage
\begin{figure}
  \centering
    \includegraphics[width=42mm,height=65mm,angle=0]{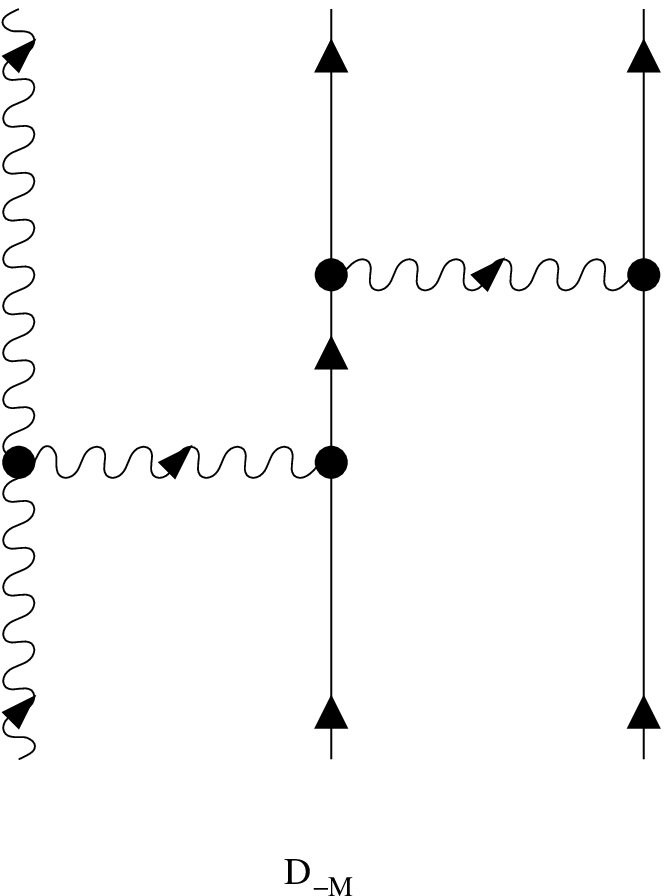}
      \hspace{1.2cm}
    \includegraphics[width=42mm,height=65mm,angle=0]{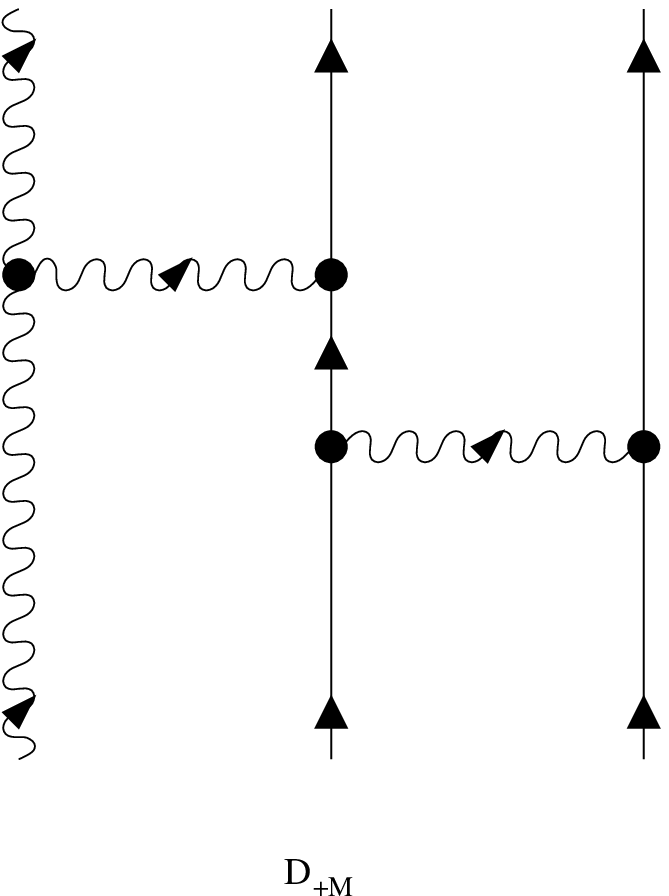}
      \hspace{1.2cm}
    \includegraphics[width=42mm,height=65mm,angle=0]{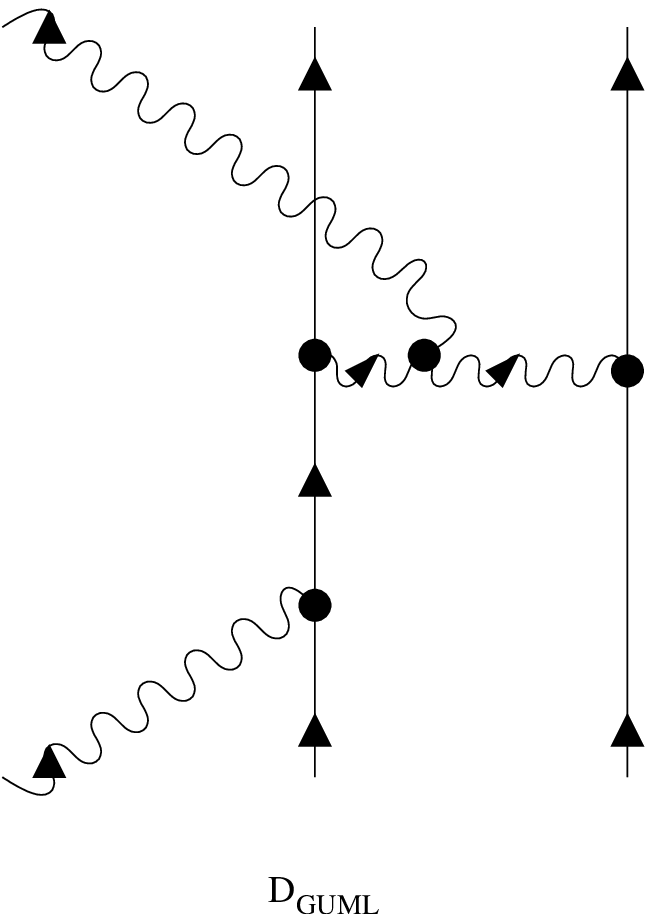}
      \vskip 26pt
    \includegraphics[width=42mm,height=65mm,angle=0]{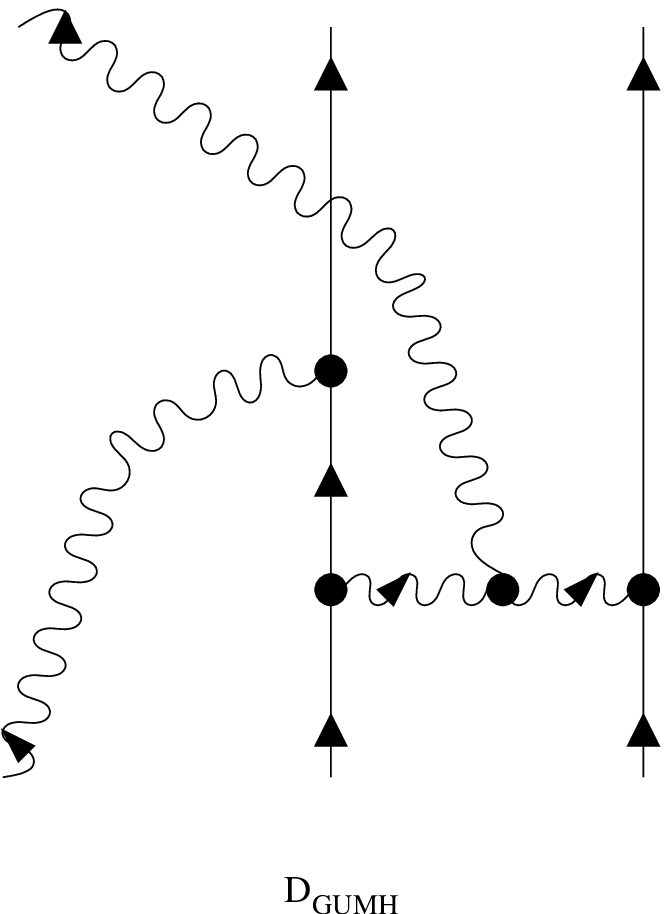}
      \hspace{1.2cm}
    \includegraphics[width=42mm,height=65mm,angle=0]{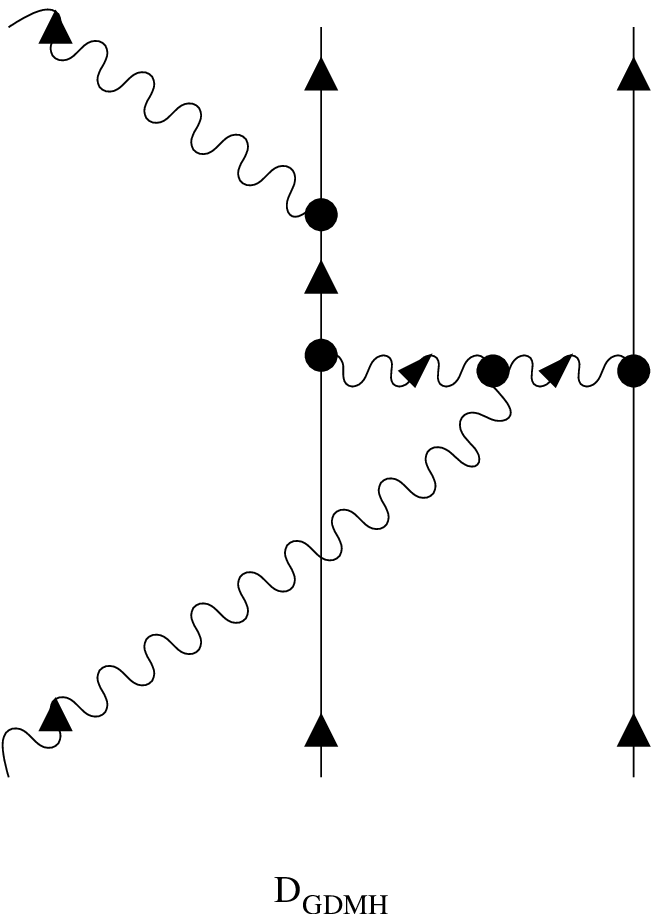}
      \hspace{1.2cm}
    \includegraphics[width=42mm,height=65mm,angle=0]{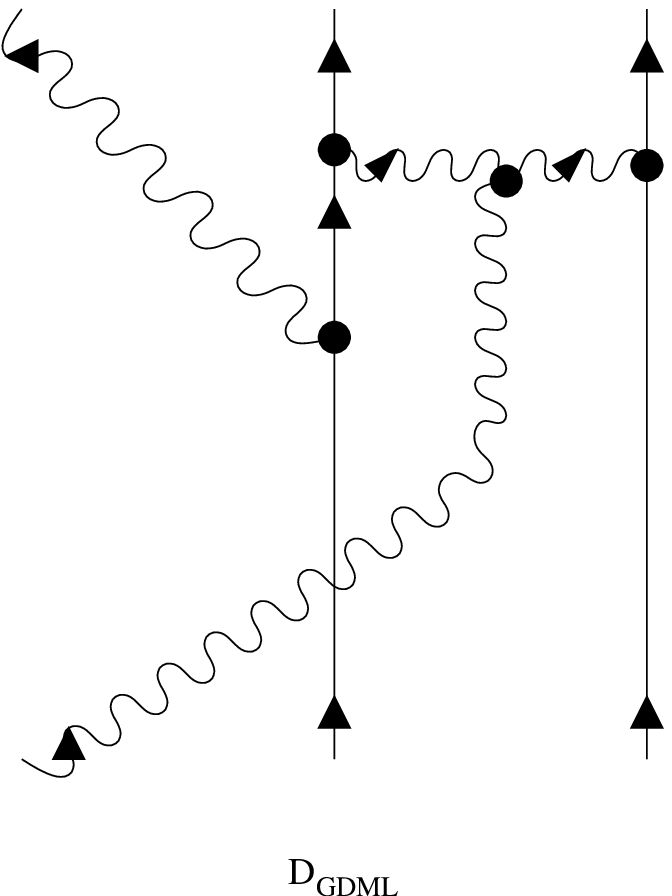}
\caption{Elastic gluon-quark-quark scattering.}
\label{fig3}
\end{figure}

\newpage
\begin{figure}
  \centering
    \includegraphics[width=42mm,height=65mm,angle=0]{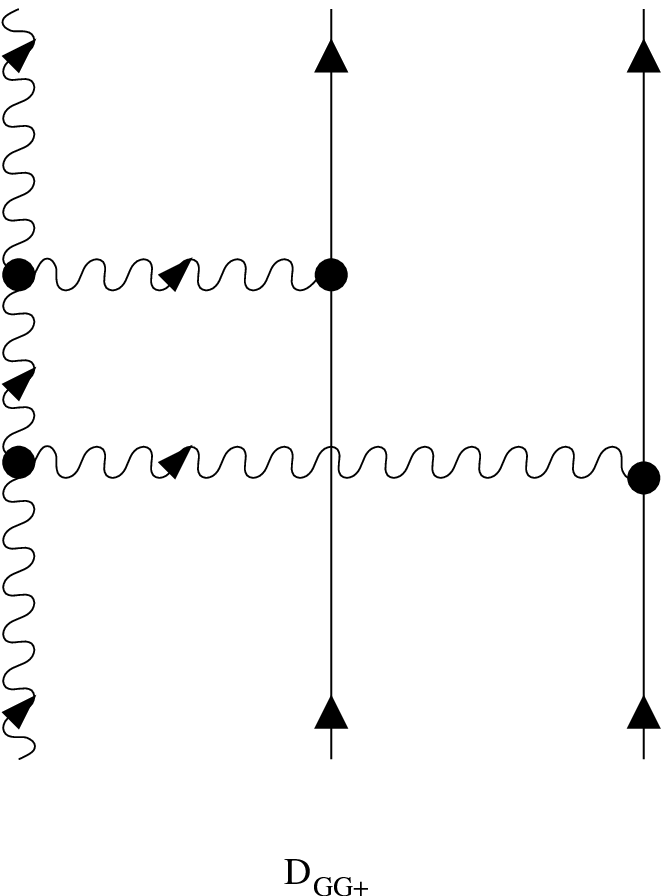}
      \hspace{1.2cm}
    \includegraphics[width=42mm,height=65mm,angle=0]{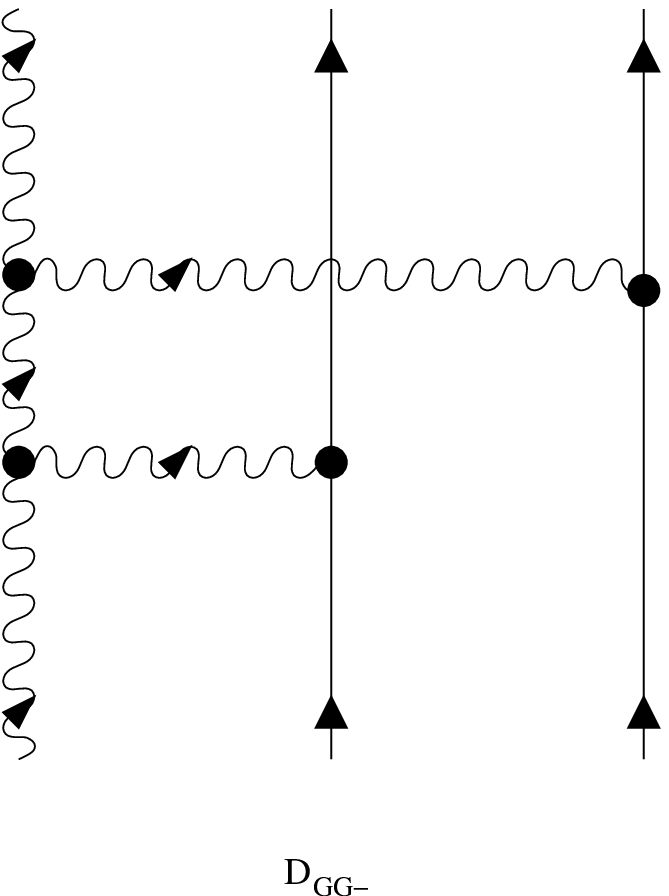}
      \vskip 26pt
    \includegraphics[width=42mm,height=65mm,angle=0]{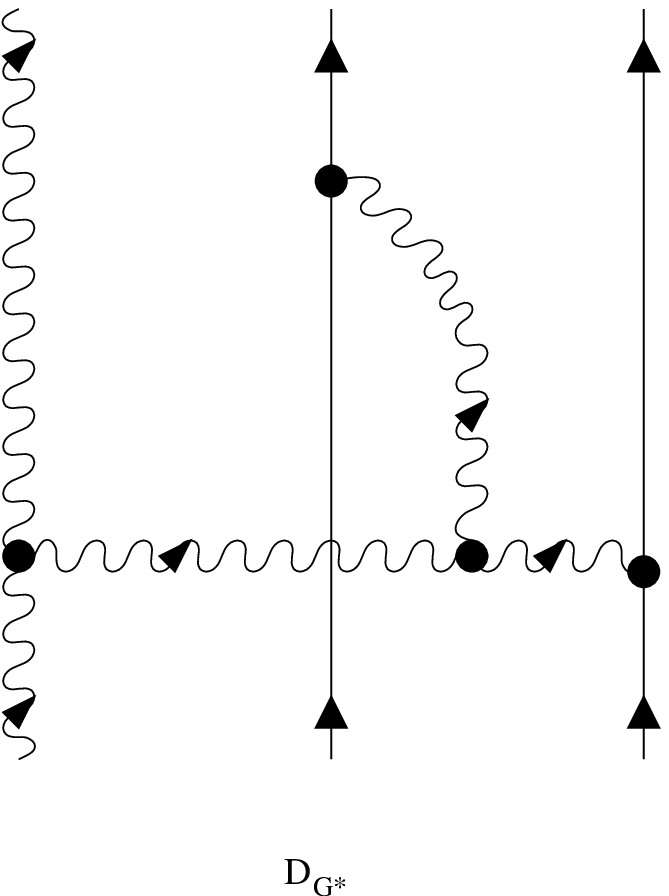}
      \hspace{1.2cm}
    \includegraphics[width=42mm,height=65mm,angle=0]{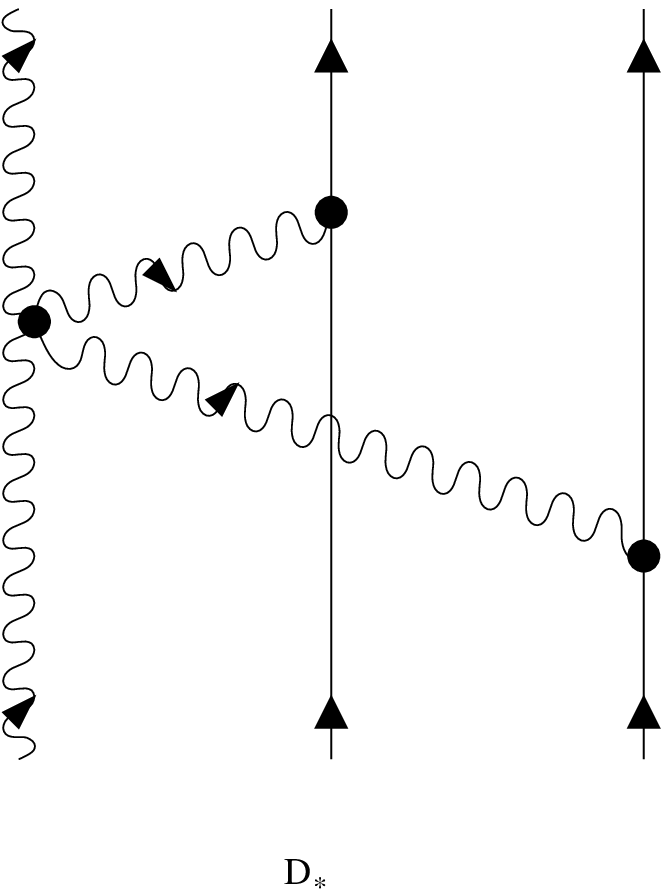}
\caption{Elastic gluon-quark-quark scattering.}
\label{fig4}
\end{figure}

\newpage
\begin{figure}
  \centering
    \includegraphics[width=42mm,height=65mm,angle=0]{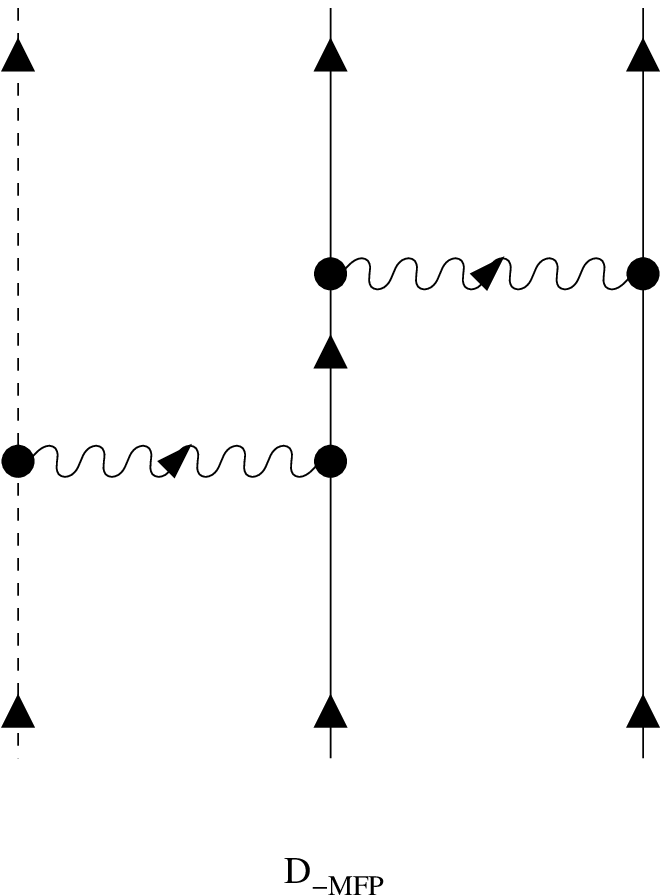}
      \hspace{1.2cm}
    \includegraphics[width=42mm,height=65mm,angle=0]{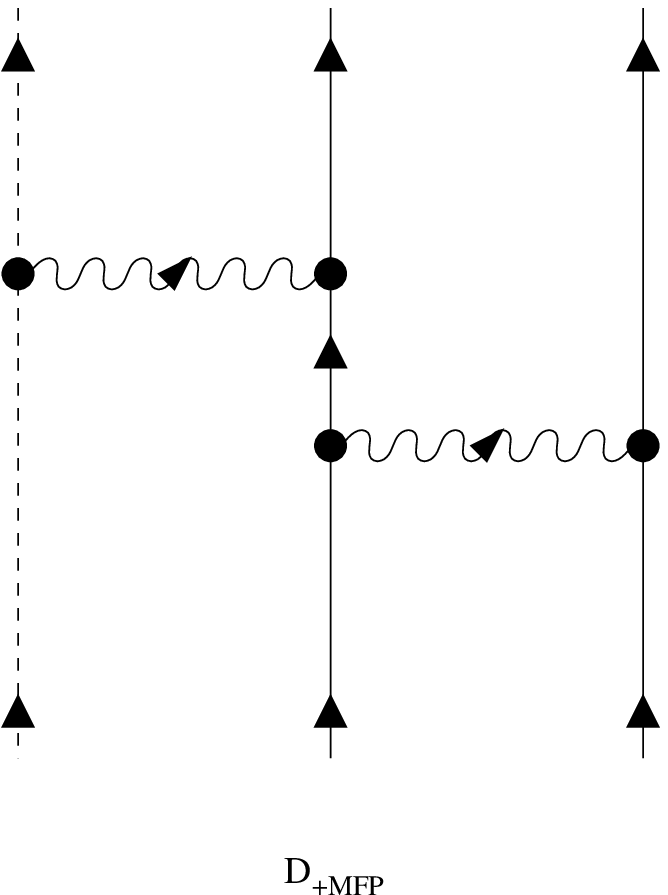}
       \vskip 26pt
    \includegraphics[width=42mm,height=65mm,angle=0]{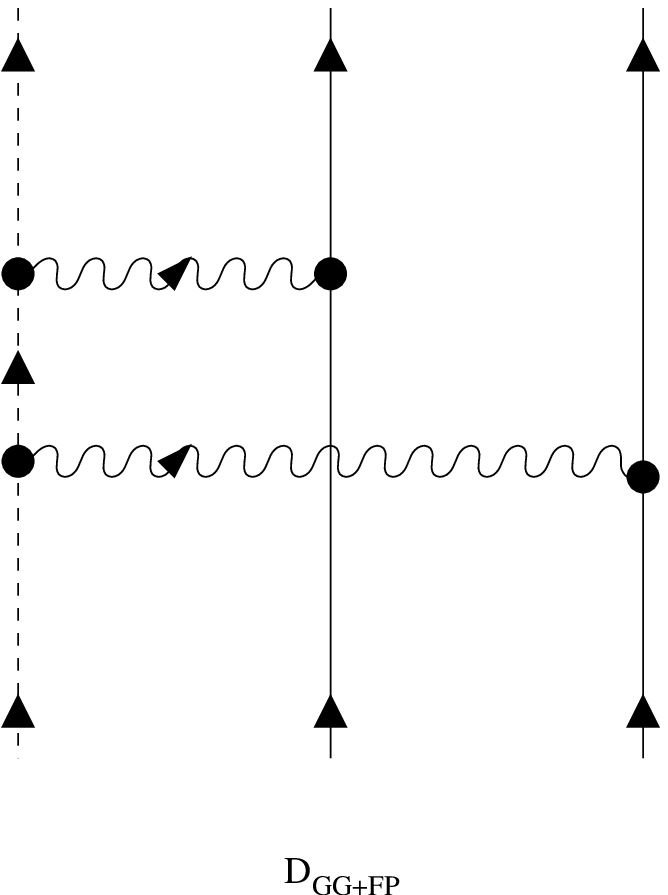}
      \hspace{1.2cm}
    \includegraphics[width=42mm,height=65mm,angle=0]{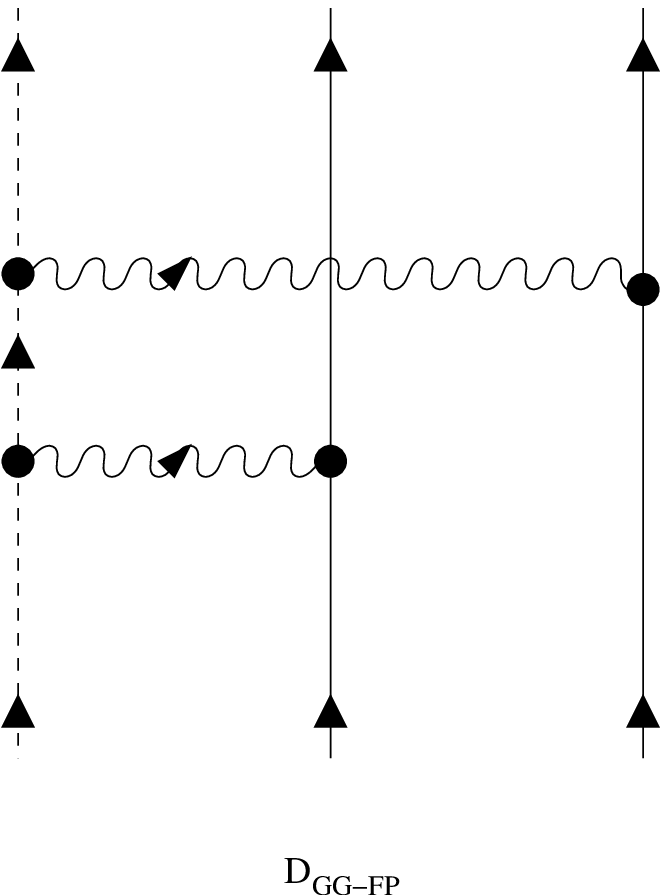}
      \hspace{1.2cm}
    \includegraphics[width=42mm,height=65mm,angle=0]{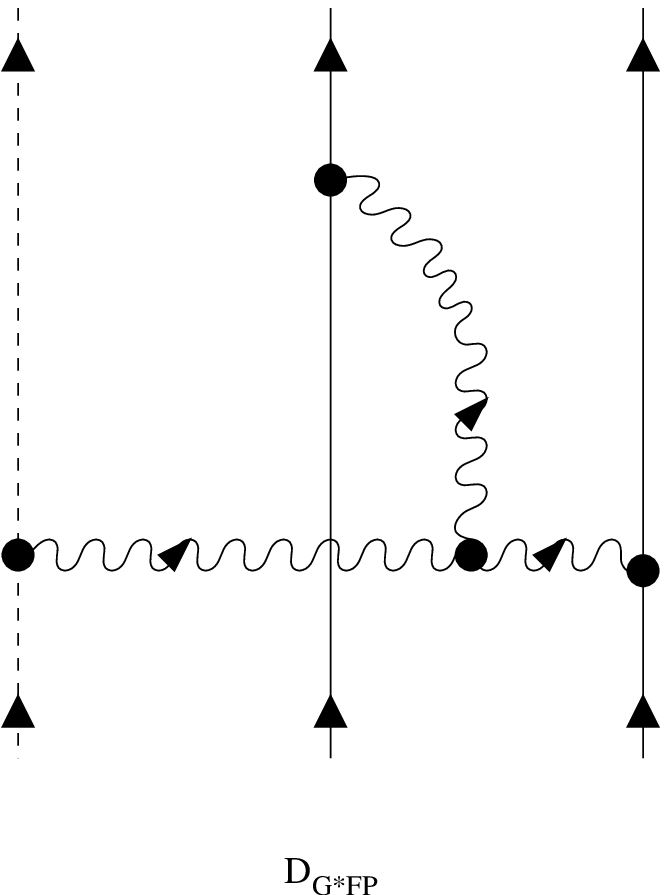}
\caption{Elastic ghost-quark-quark scattering.}
\label{fig5}
\end{figure}

\newpage
\begin{figure}
  \centering
    \includegraphics[width=42mm,height=65mm,angle=0]{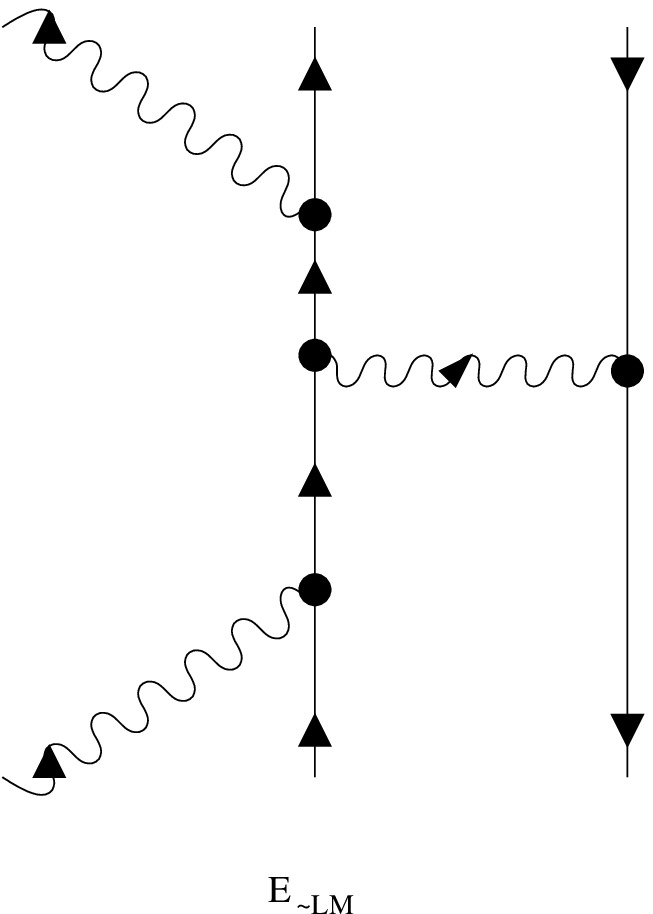}
      \hspace{1.2cm}
    \includegraphics[width=42mm,height=65mm,angle=0]{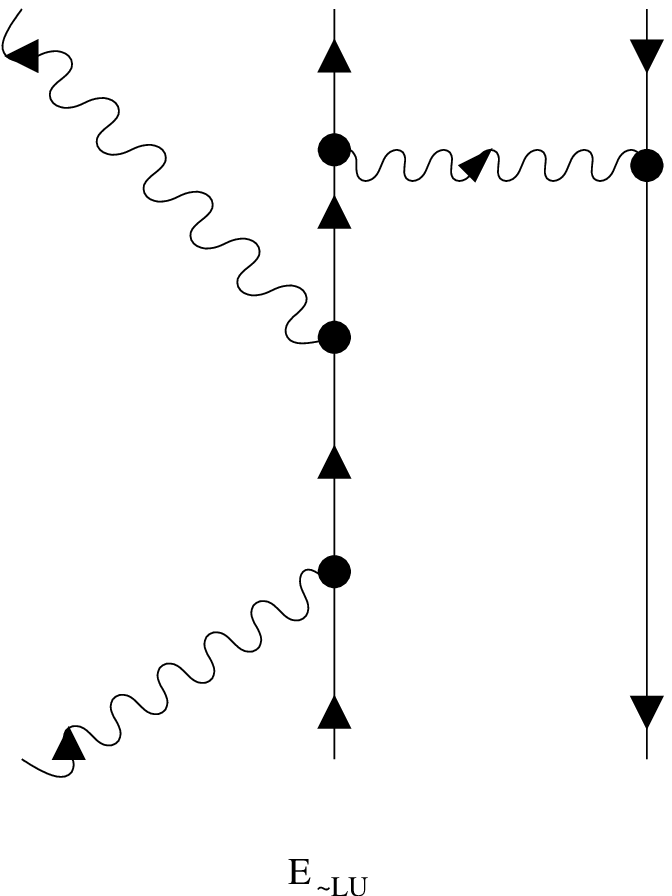}
      \hspace{1.2cm}
    \includegraphics[width=42mm,height=65mm,angle=0]{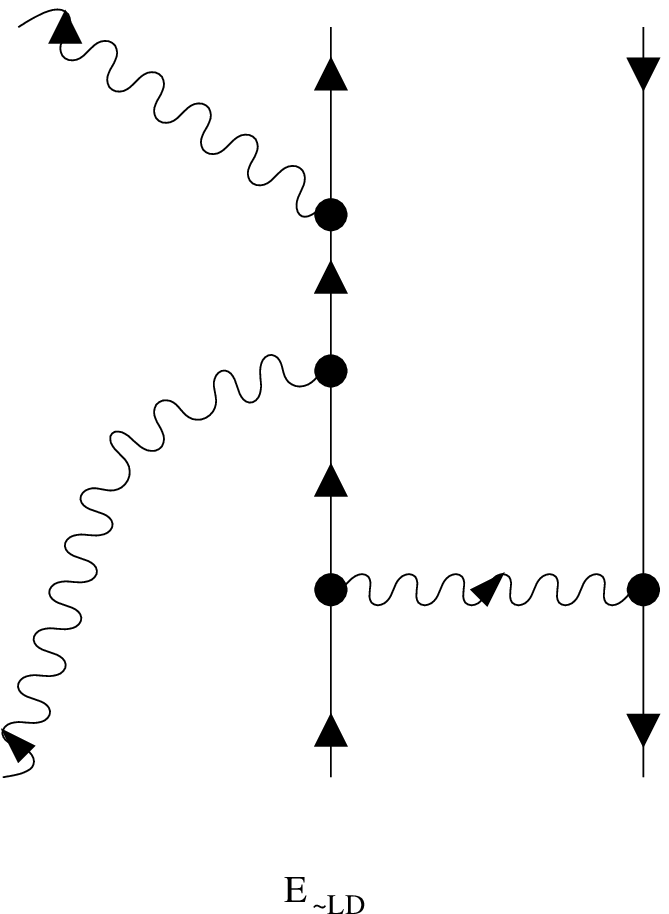}
      \vskip 26pt
    \includegraphics[width=42mm,height=65mm,angle=0]{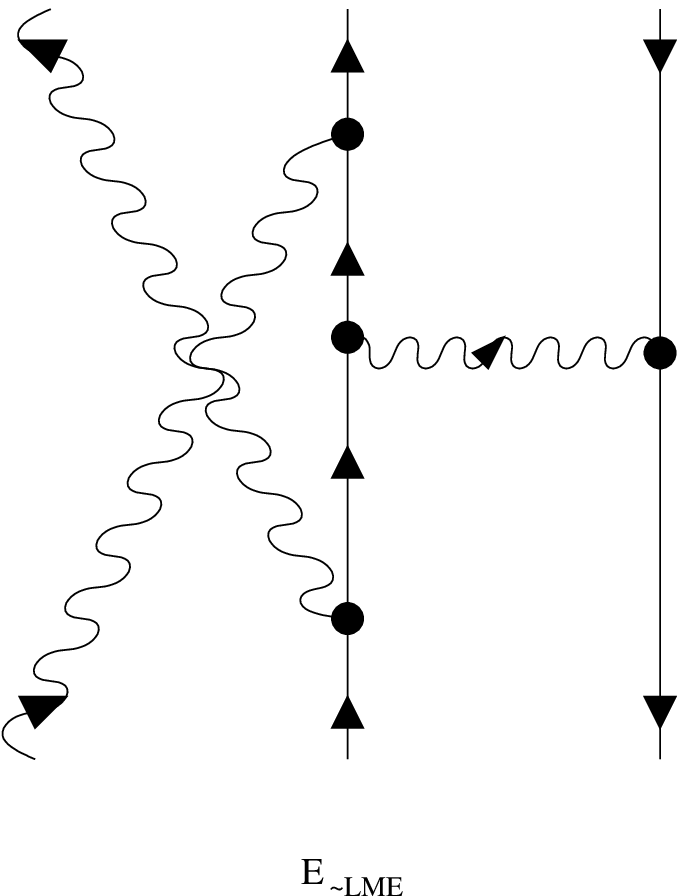}
      \hspace{1.2cm}
    \includegraphics[width=42mm,height=65mm,angle=0]{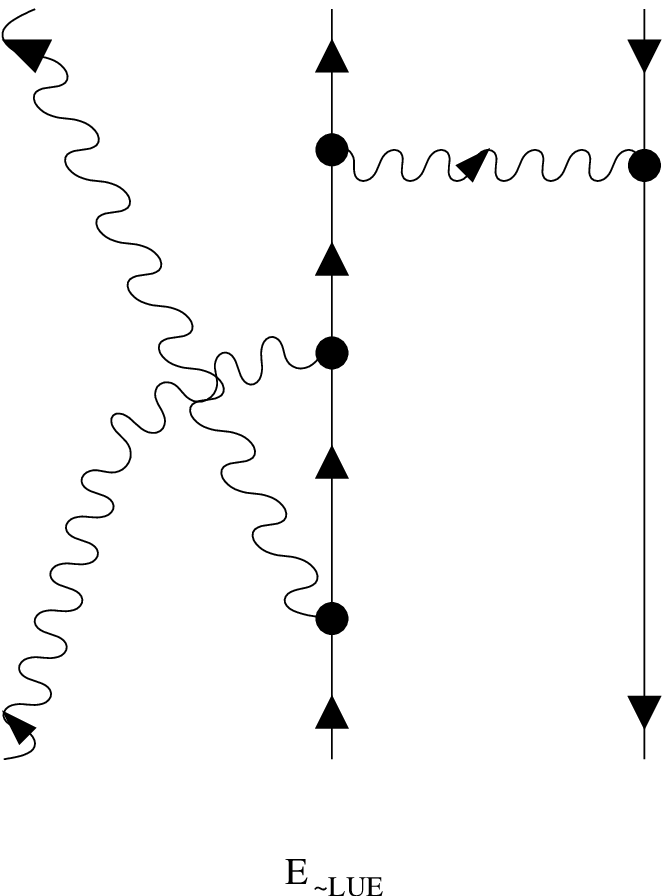}
      \hspace{1.2cm}
    \includegraphics[width=42mm,height=65mm,angle=0]{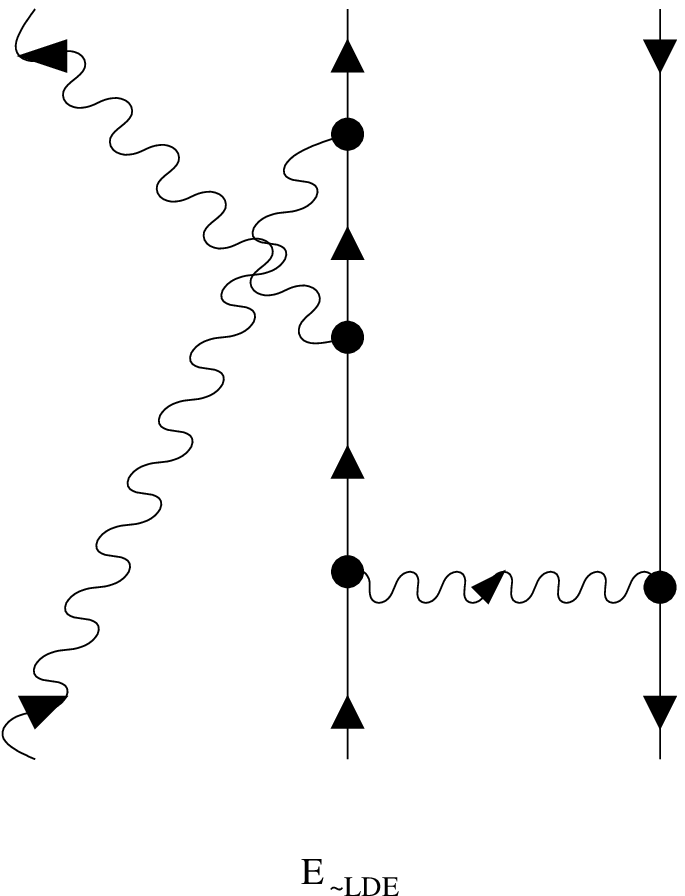}
\caption{Elastic gluon-quark-antiquark scattering.}
\label{fig6}
\end{figure}

\newpage
\begin{figure}
  \centering
    \includegraphics[width=42mm,height=65mm,angle=0]{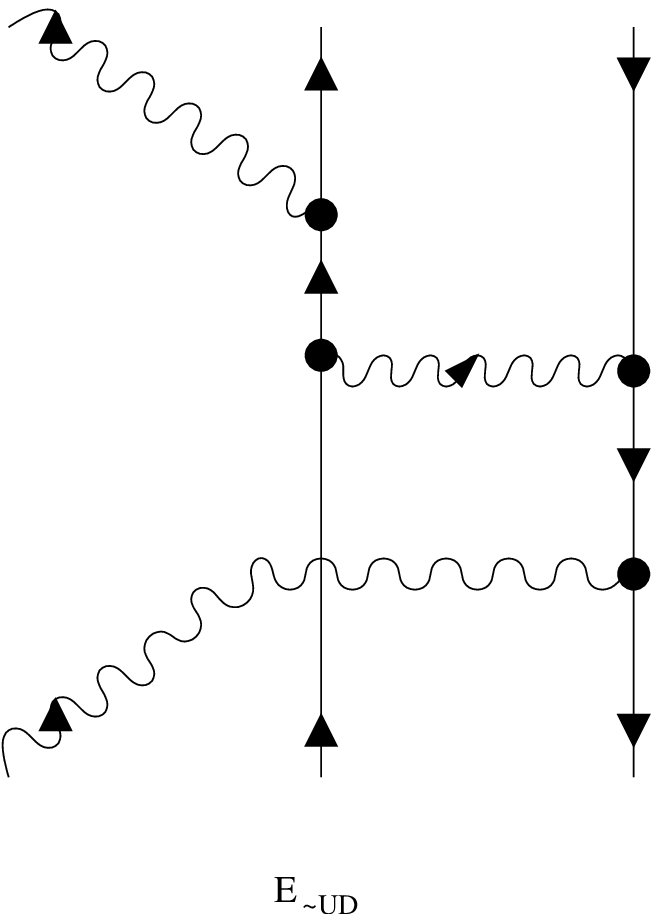}
      \hspace{1.2cm}
    \includegraphics[width=42mm,height=65mm,angle=0]{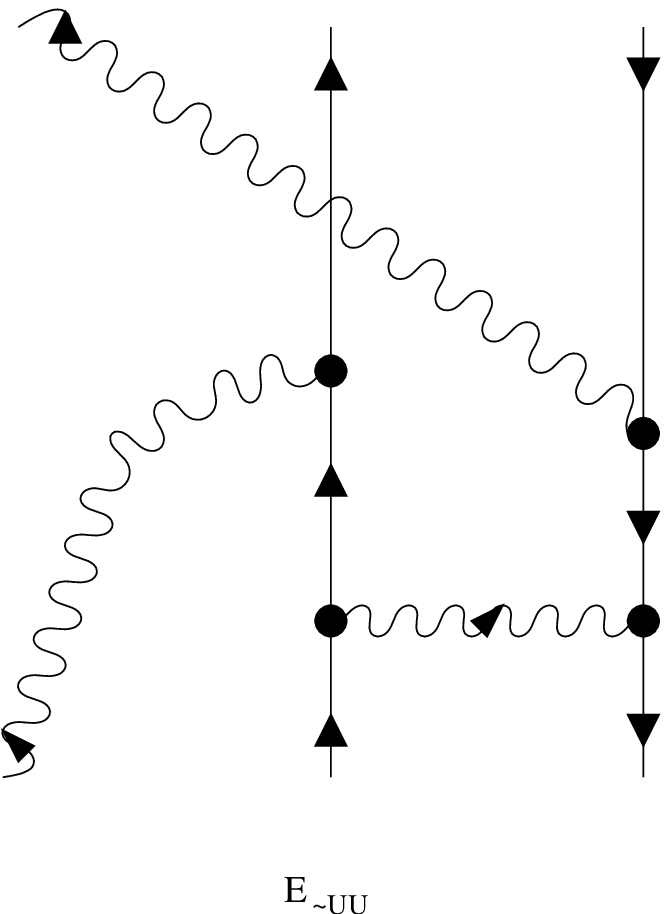}
      \vskip 26pt
    \includegraphics[width=42mm,height=65mm,angle=0]{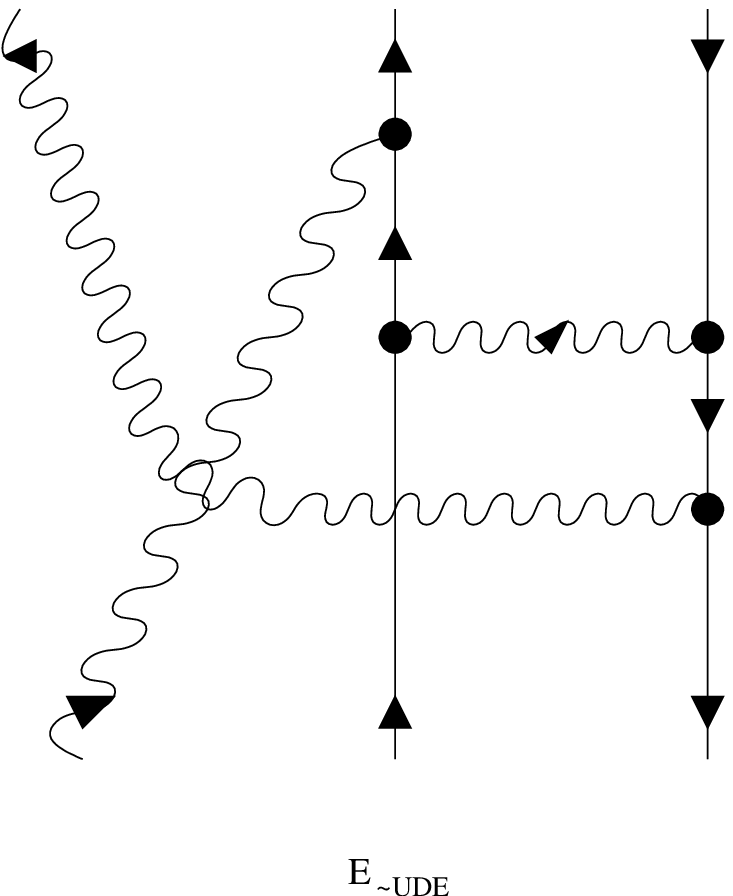}
      \hspace{1.2cm}
    \includegraphics[width=42mm,height=65mm,angle=0]{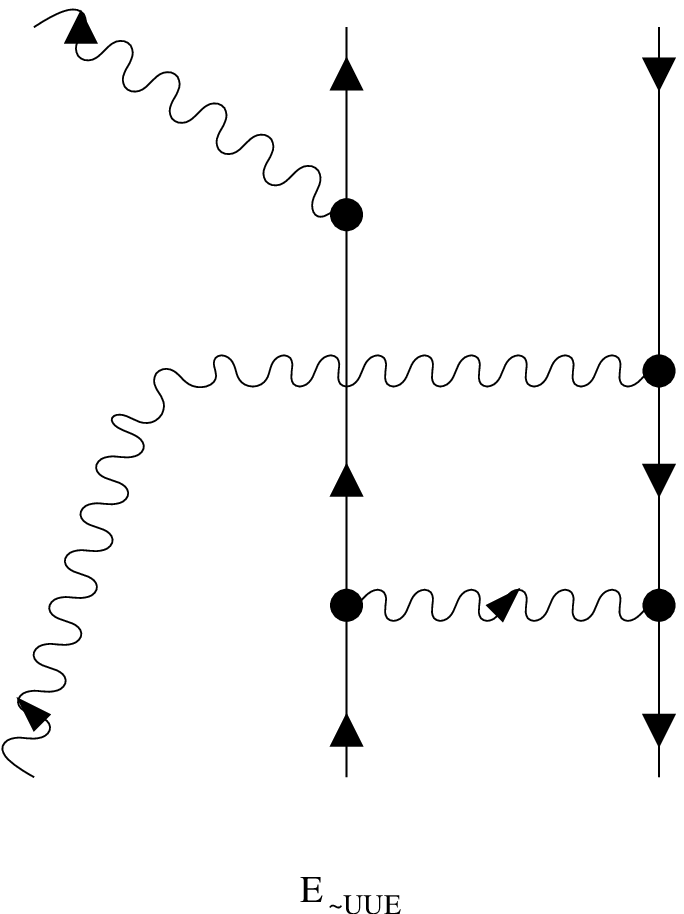}
\caption{Elastic gluon-quark-antiquark scattering.}
\label{fig7}
\end{figure}

\newpage
\begin{figure}
  \centering
    \includegraphics[width=42mm,height=65mm,angle=0]{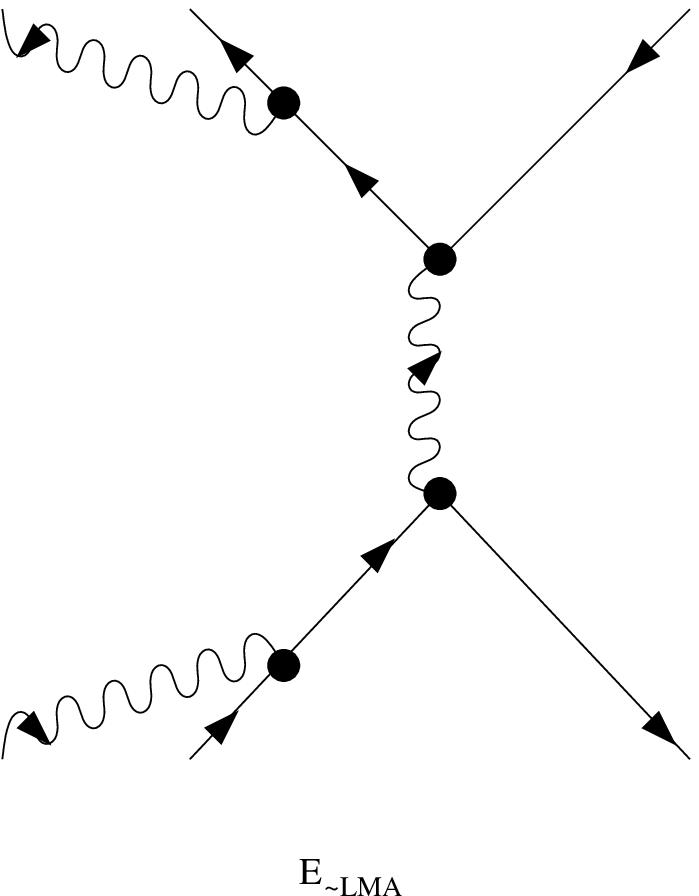}
      \hspace{1.2cm}
    \includegraphics[width=42mm,height=65mm,angle=0]{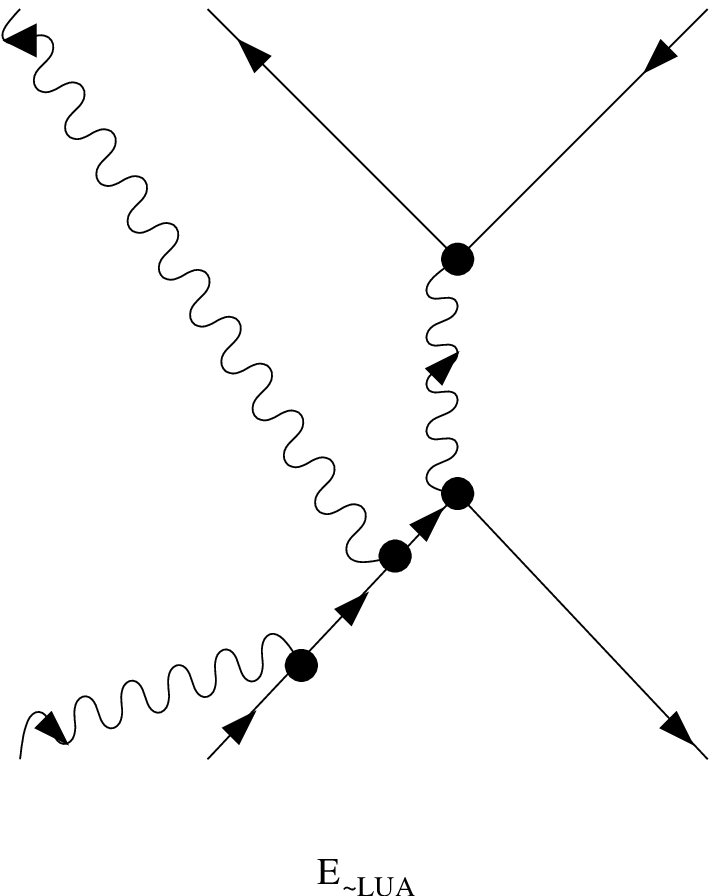}
      \hspace{1.2cm}
    \includegraphics[width=42mm,height=65mm,angle=0]{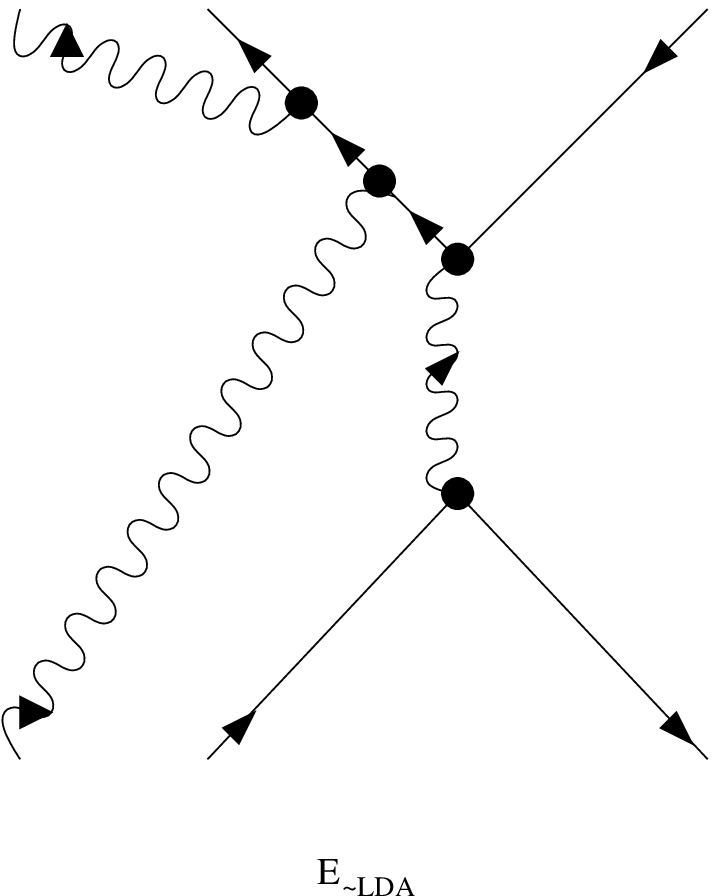}
      \vskip 26pt
    \includegraphics[width=42mm,height=65mm,angle=0]{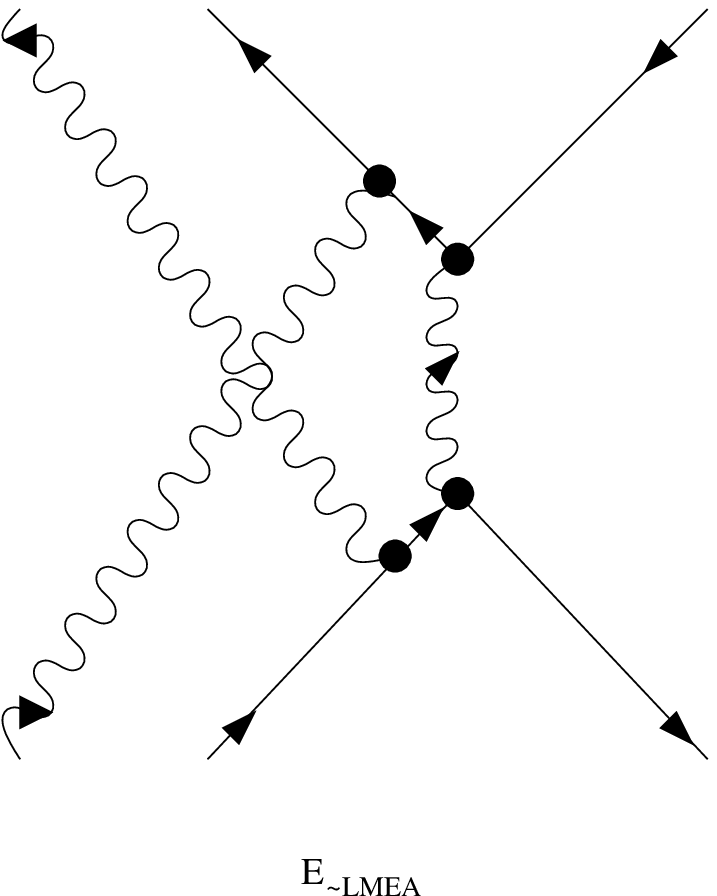}
      \hspace{1.2cm}
    \includegraphics[width=42mm,height=65mm,angle=0]{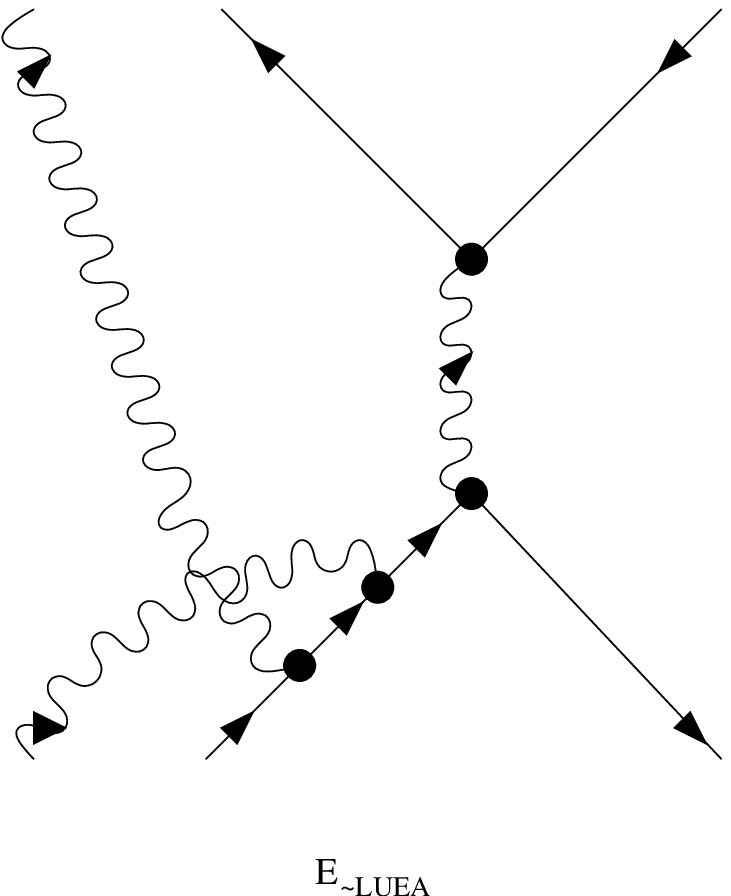}
      \hspace{1.2cm}
    \includegraphics[width=42mm,height=65mm,angle=0]{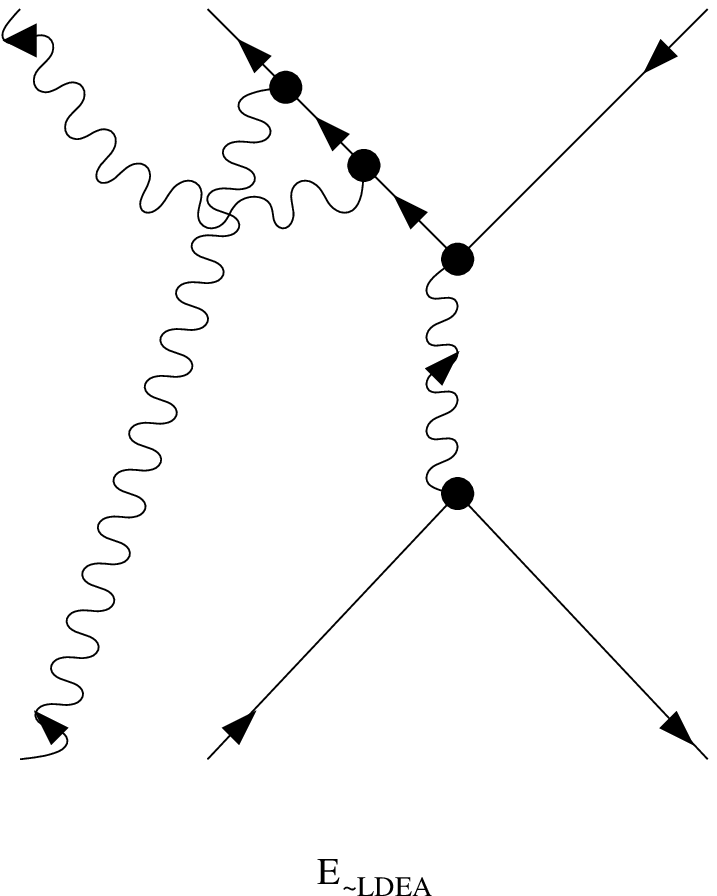}
\caption{Elastic gluon-quark-antiquark scattering.}
\label{fig8}
\end{figure}

\newpage
\begin{figure}
  \centering
    \includegraphics[width=42mm,height=65mm,angle=0]{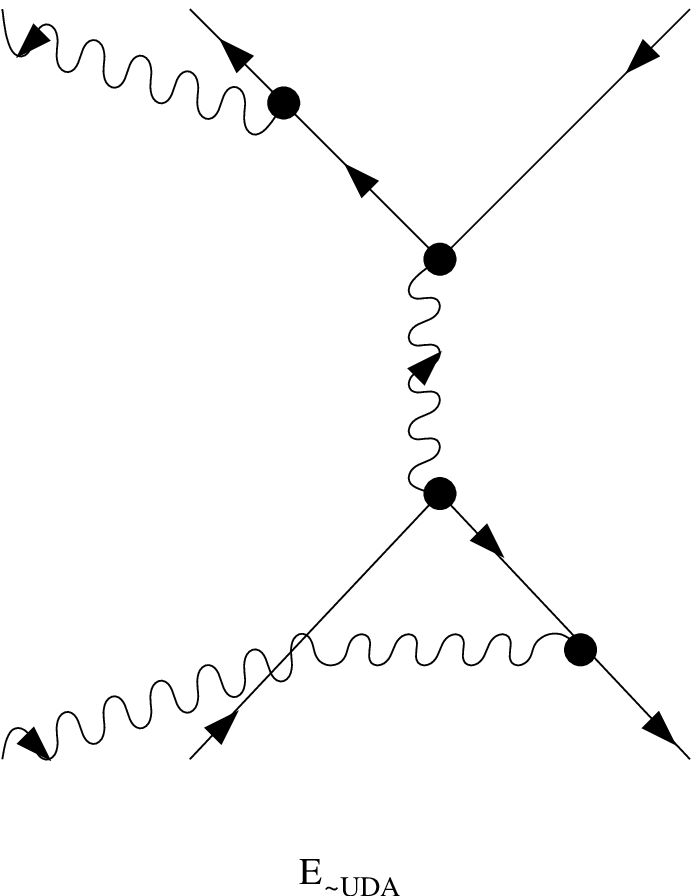}
      \hspace{1.2cm}
    \includegraphics[width=42mm,height=65mm,angle=0]{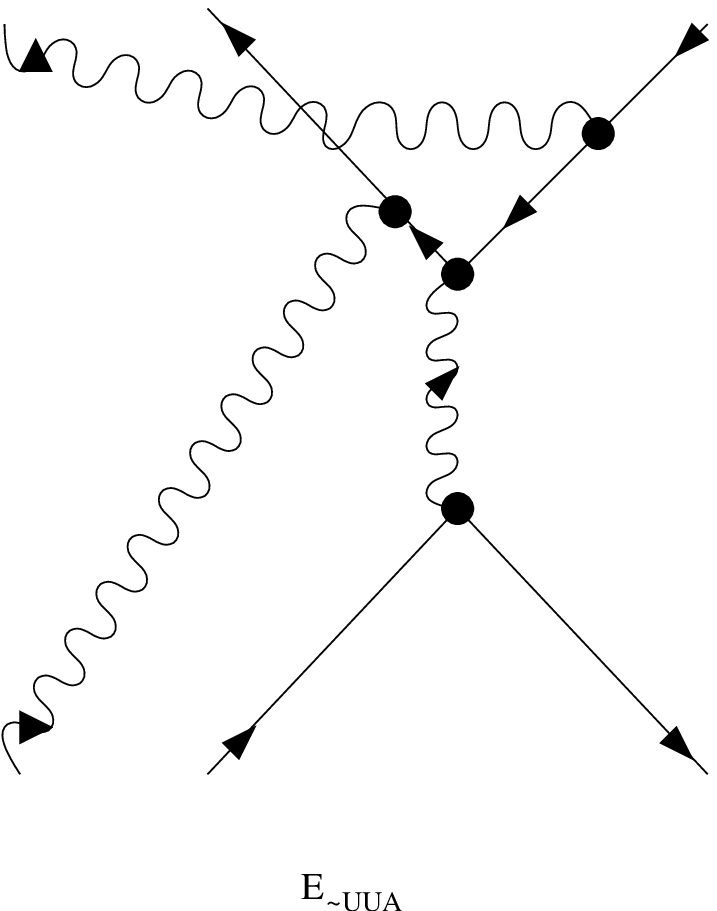}
      \vskip 26pt
    \includegraphics[width=42mm,height=65mm,angle=0]{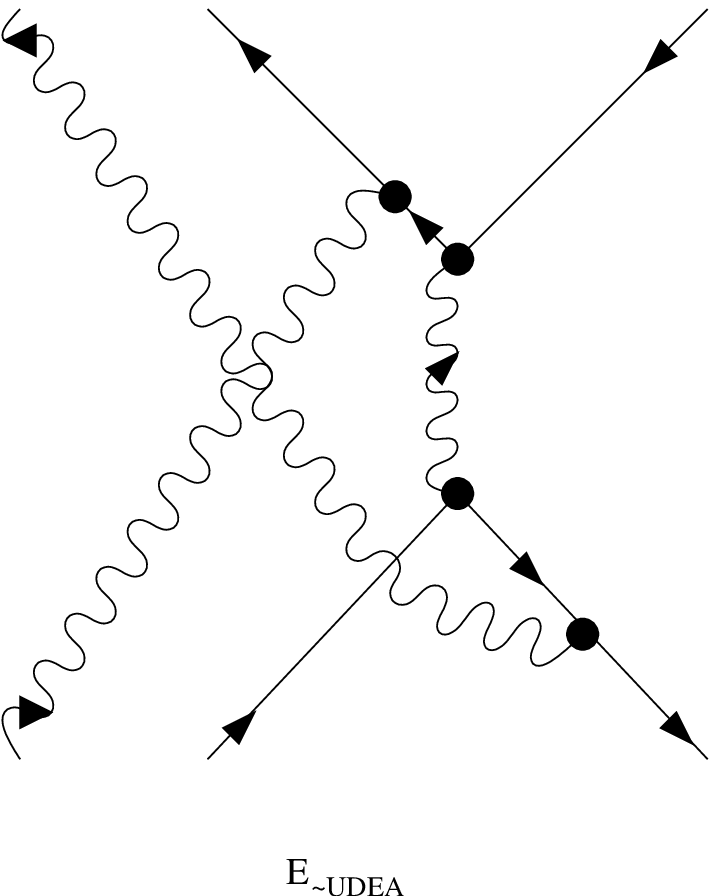}
      \hspace{1.2cm}
    \includegraphics[width=42mm,height=65mm,angle=0]{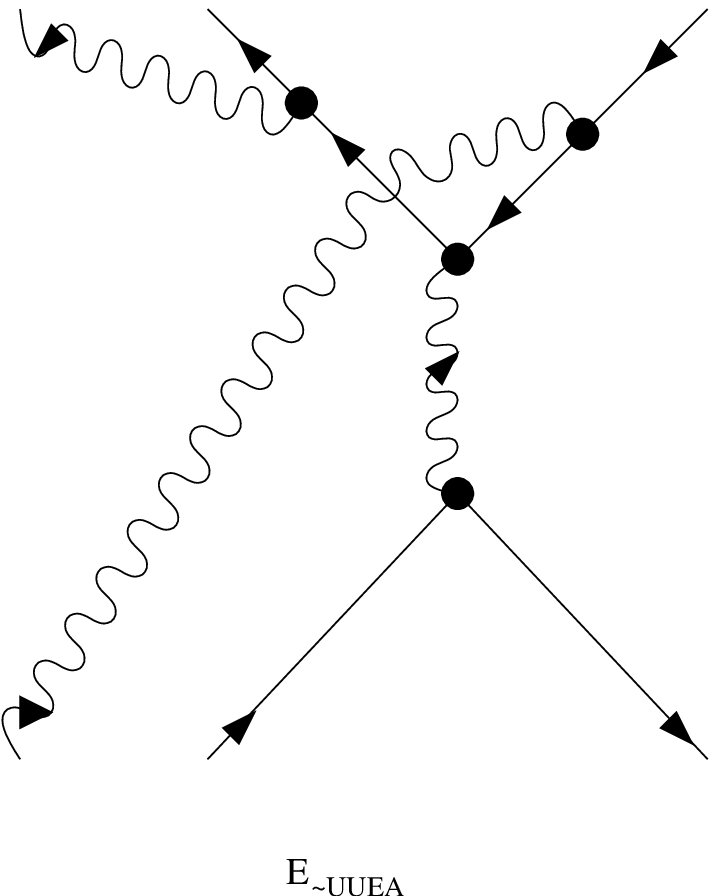}
\caption{Elastic gluon-quark-antiquark scattering.}
\label{fig9}
\end{figure}

\newpage
\begin{figure}
  \centering
    \includegraphics[width=42mm,height=65mm,angle=0]{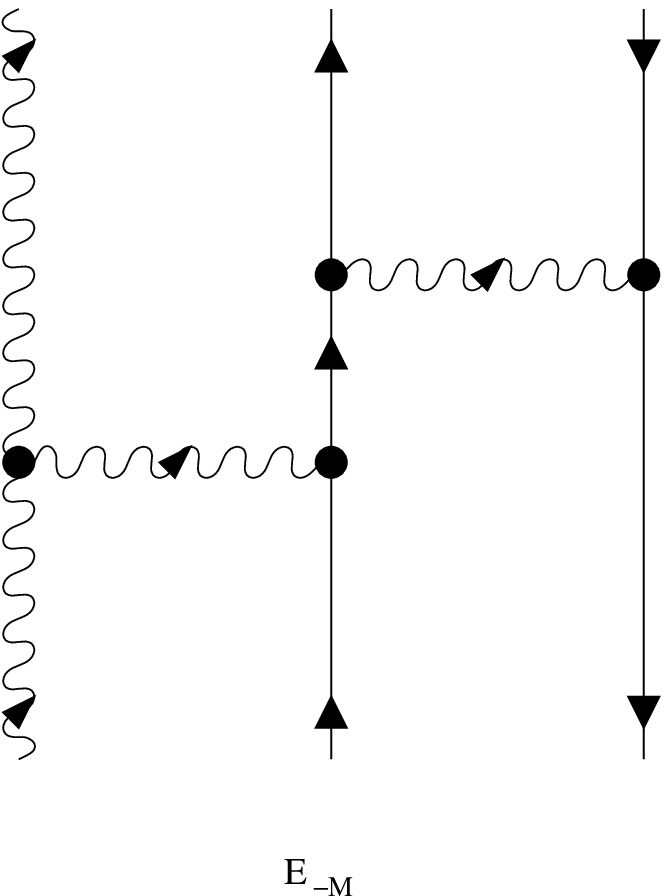}
      \hspace{1.2cm}
    \includegraphics[width=42mm,height=65mm,angle=0]{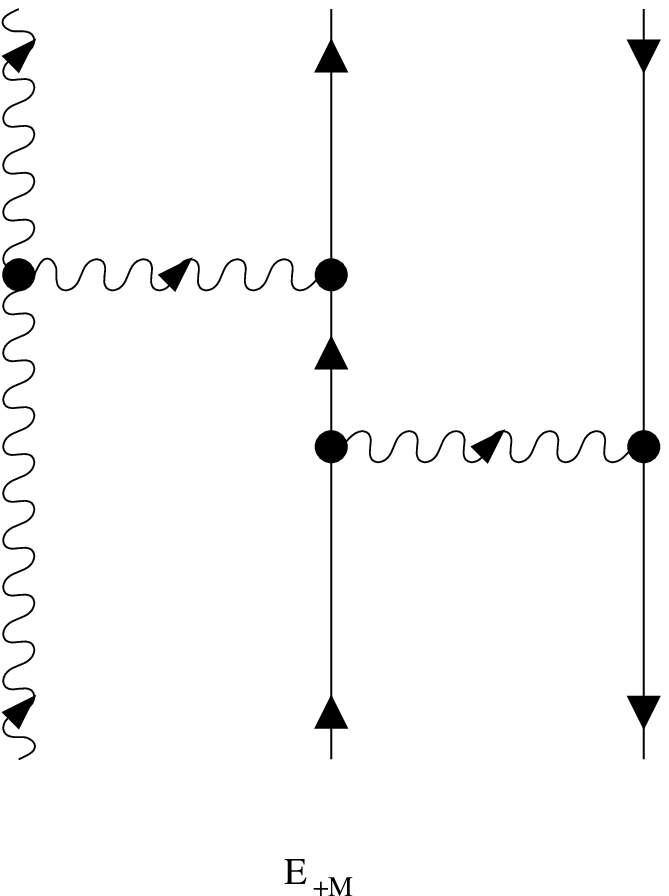}
      \hspace{1.2cm}
    \includegraphics[width=42mm,height=65mm,angle=0]{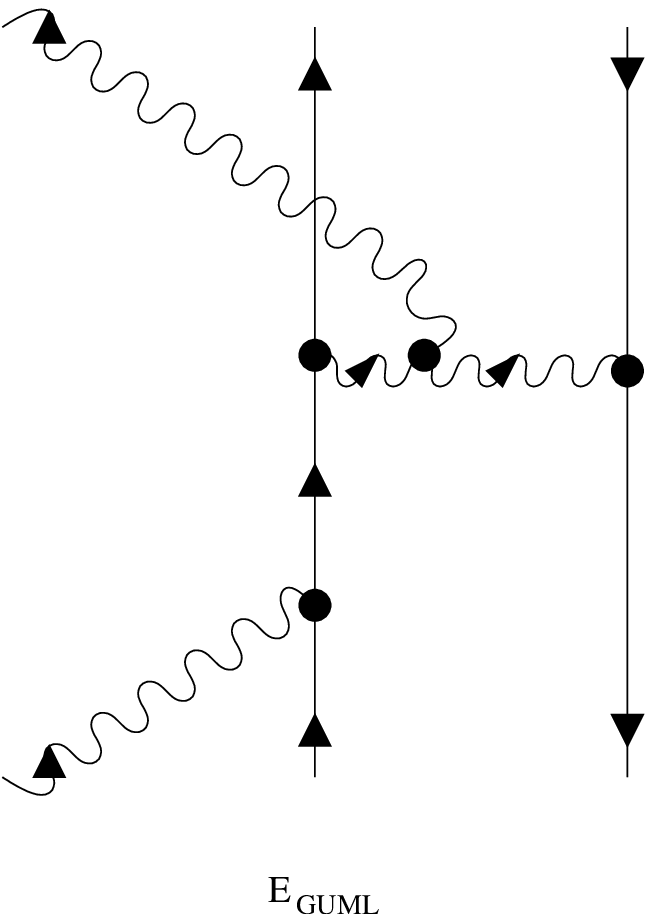}
      \vskip 26pt
    \includegraphics[width=42mm,height=65mm,angle=0]{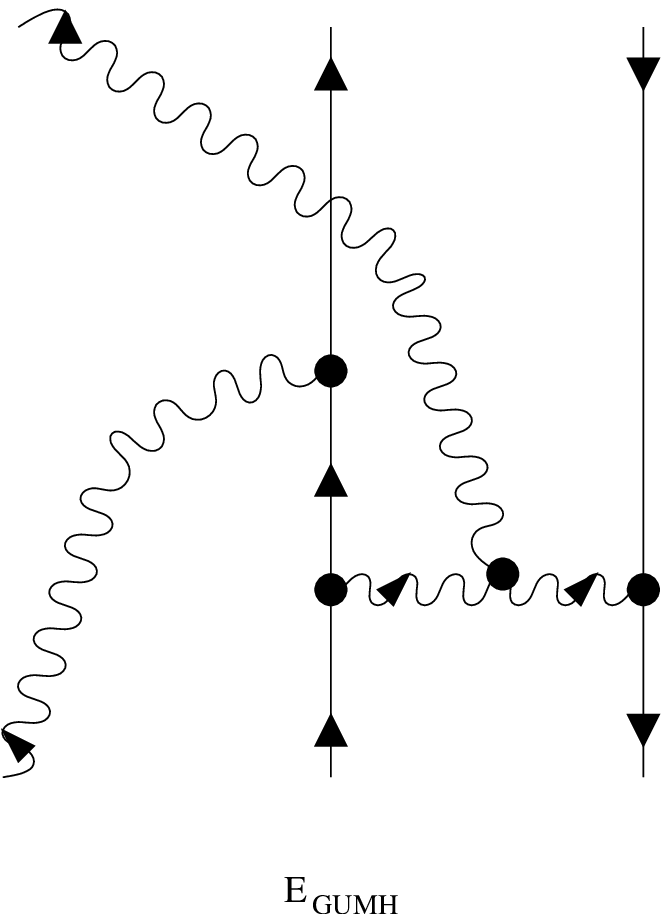}
      \hspace{1.2cm}
    \includegraphics[width=42mm,height=65mm,angle=0]{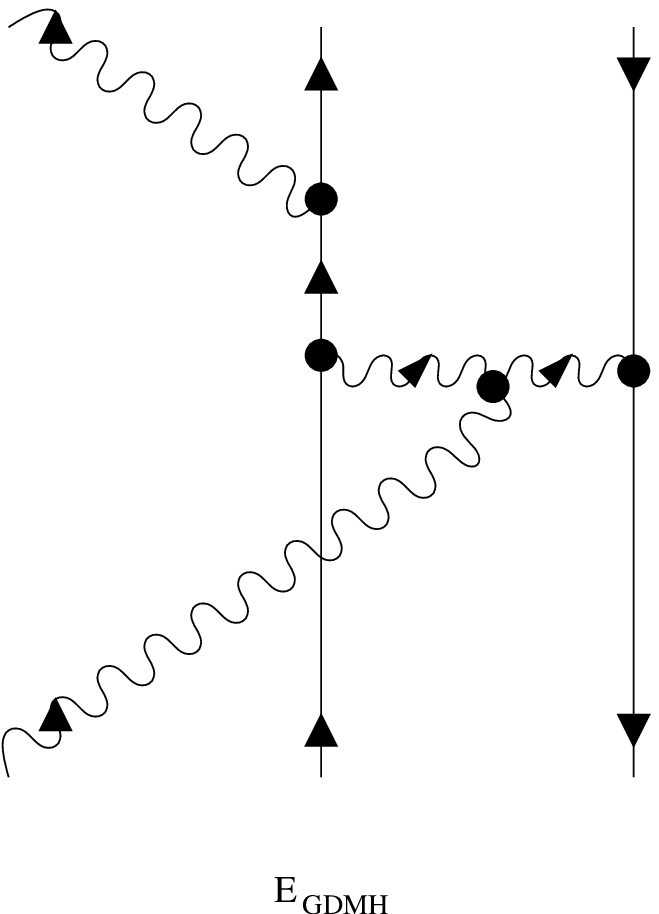}
      \hspace{1.2cm}
    \includegraphics[width=42mm,height=65mm,angle=0]{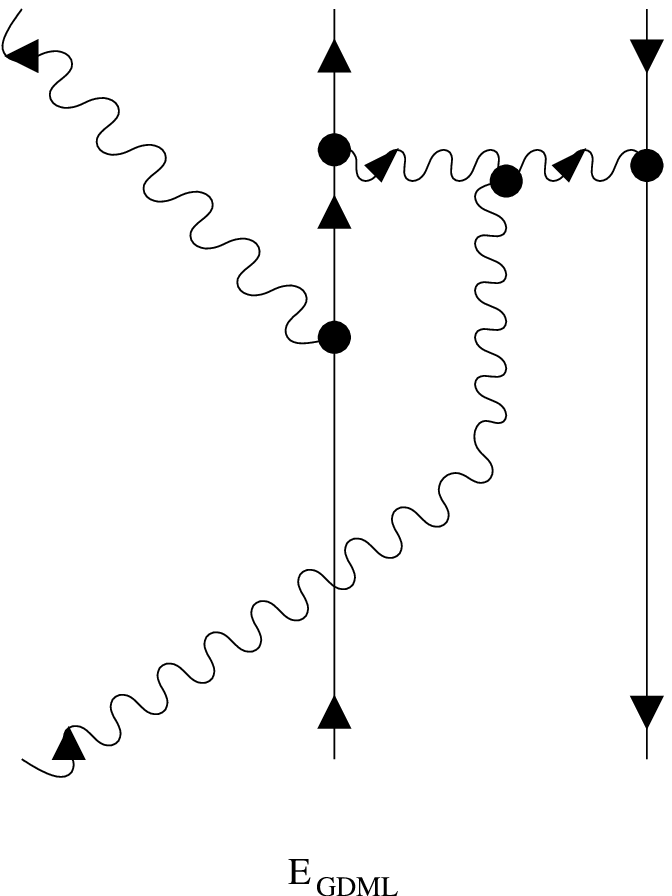}
\caption{Elastic gluon-quark-antiquark scattering.}
\label{fig10}
\end{figure}

\newpage
\begin{figure}
  \centering
    \includegraphics[width=42mm,height=65mm,angle=0]{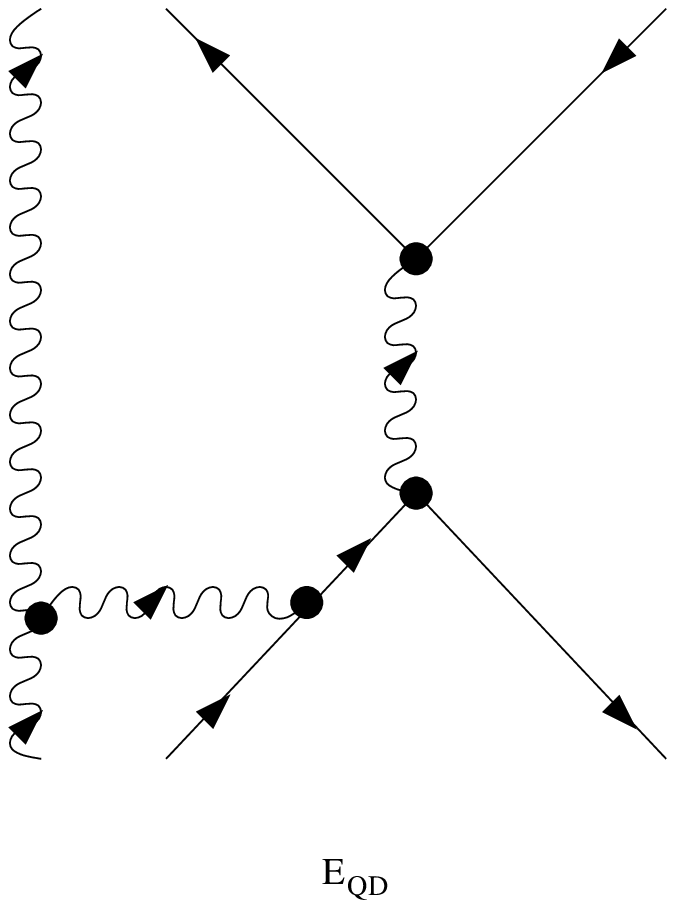}
      \hspace{1.2cm}
    \includegraphics[width=42mm,height=65mm,angle=0]{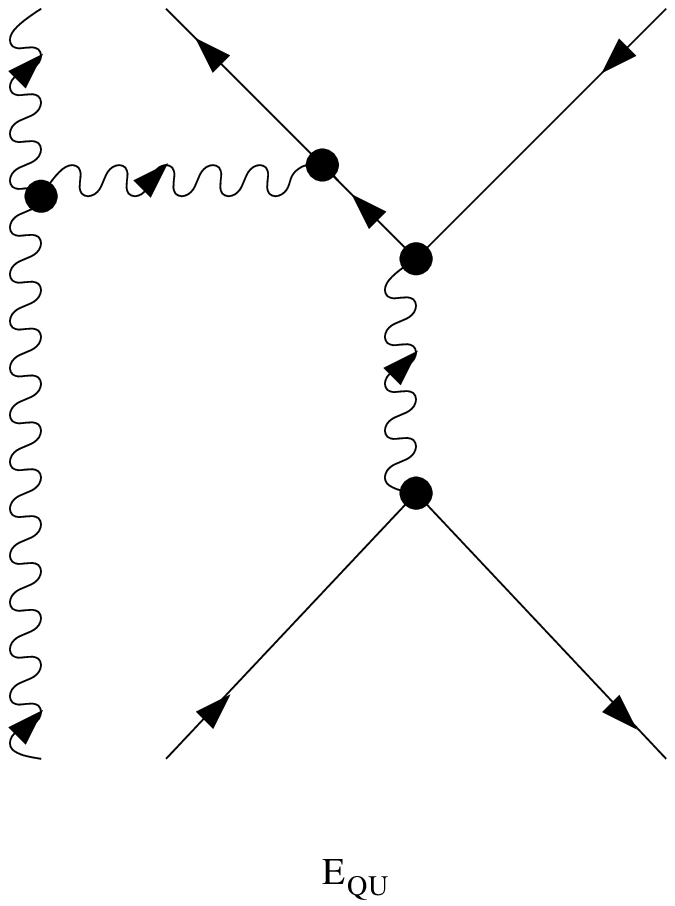}
      \hspace{1.2cm}
    \includegraphics[width=42mm,height=65mm,angle=0]{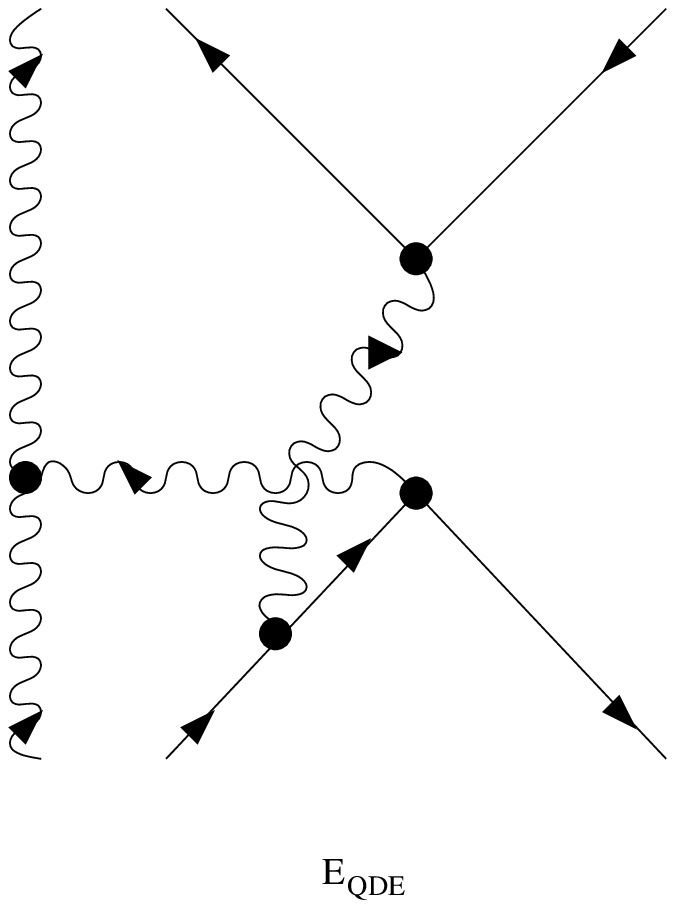}
      \vskip 26pt
    \includegraphics[width=42mm,height=65mm,angle=0]{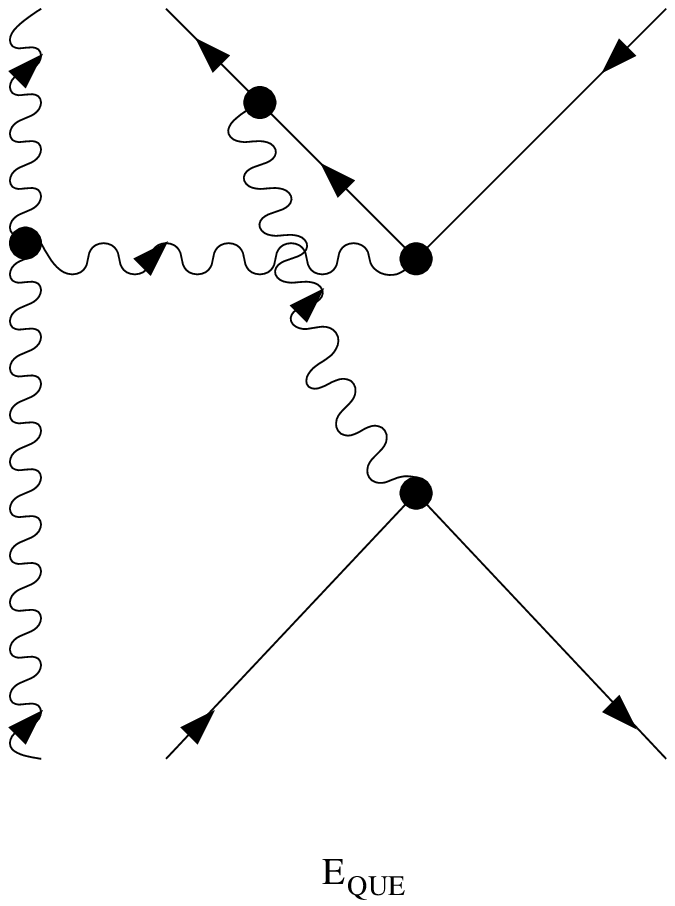}
      \hspace{1.2cm}
    \includegraphics[width=42mm,height=65mm,angle=0]{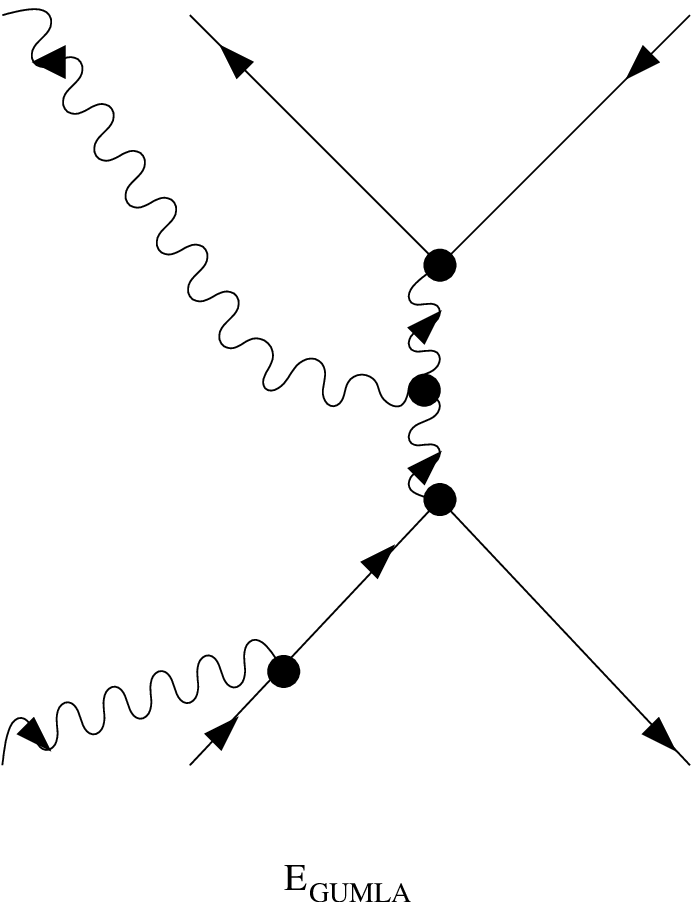}
      \hspace{1.2cm}
    \includegraphics[width=42mm,height=65mm,angle=0]{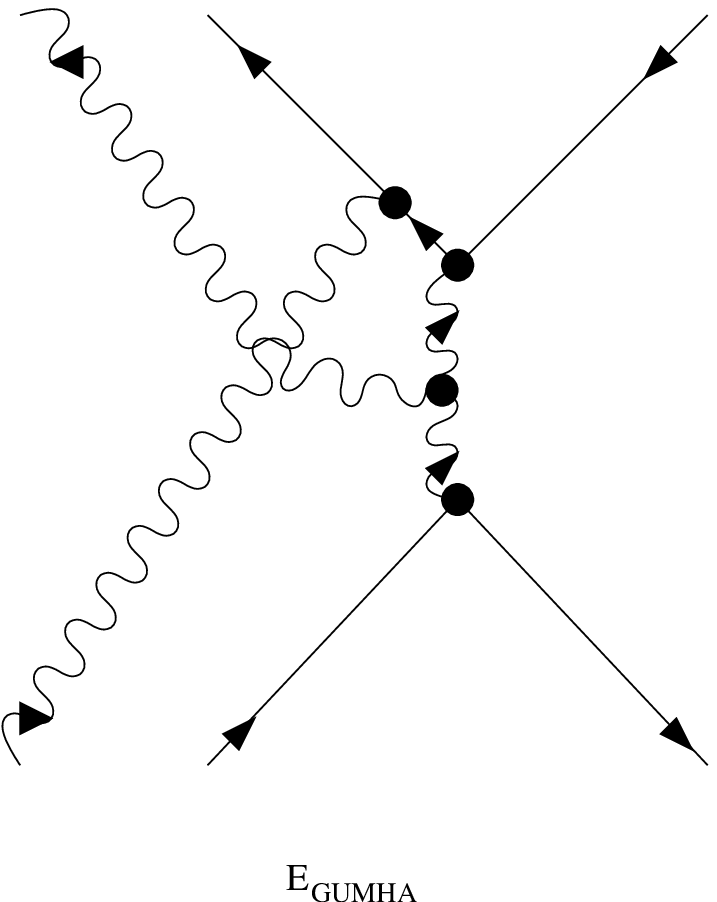}
      \vskip 26pt
    \includegraphics[width=42mm,height=65mm,angle=0]{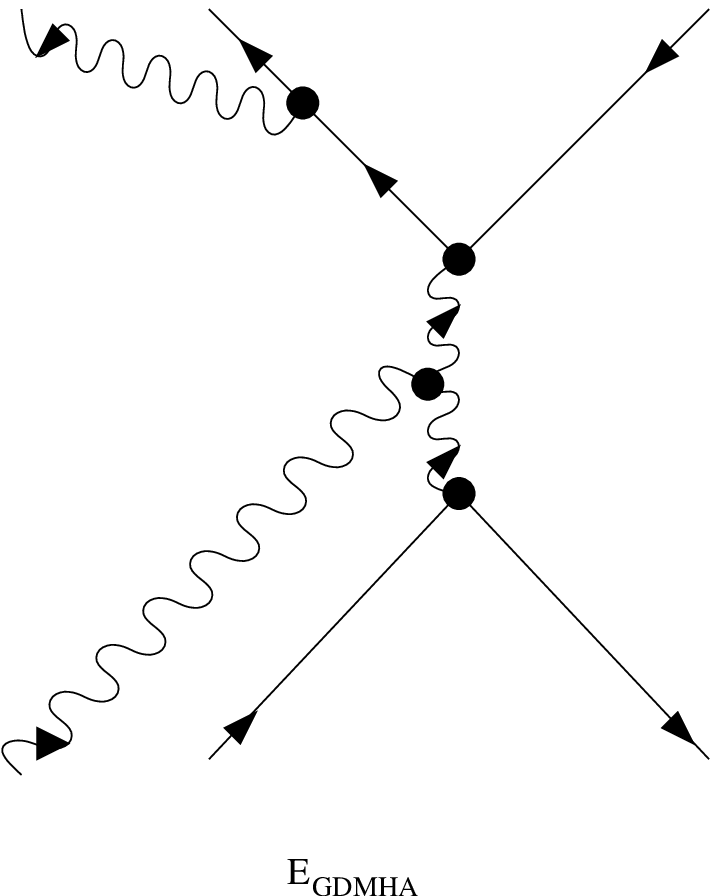}
      \hspace{1.2cm}
    \includegraphics[width=42mm,height=65mm,angle=0]{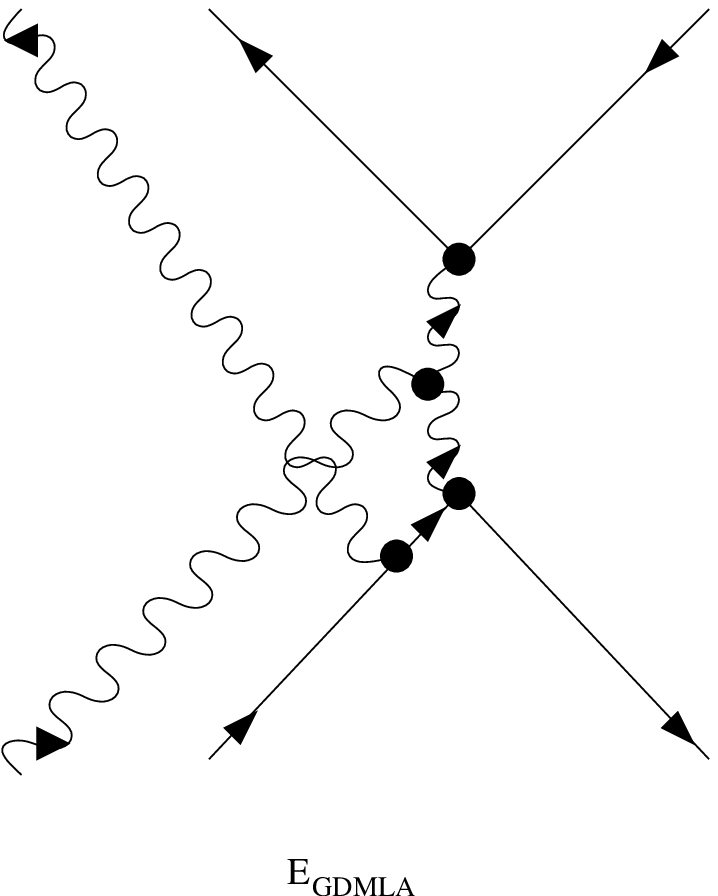}
\caption{Elastic gluon-quark-antiquark scattering.}
\label{fig11}
\end{figure}

\newpage
\begin{figure}
  \centering
    \includegraphics[width=42mm,height=65mm,angle=0]{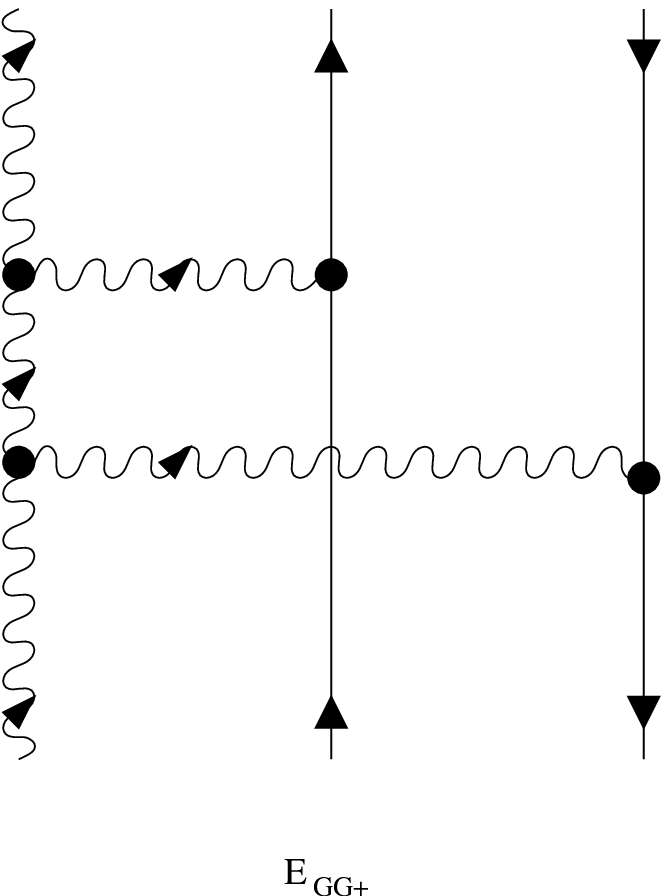}
      \hspace{1.2cm}
    \includegraphics[width=42mm,height=65mm,angle=0]{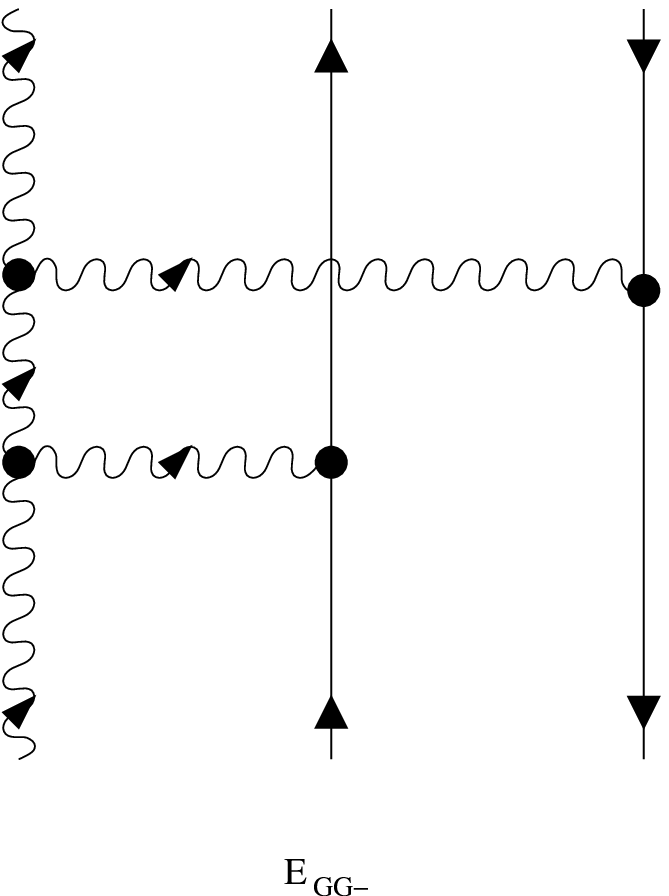}
      \hspace{1.2cm}
    \includegraphics[width=42mm,height=65mm,angle=0]{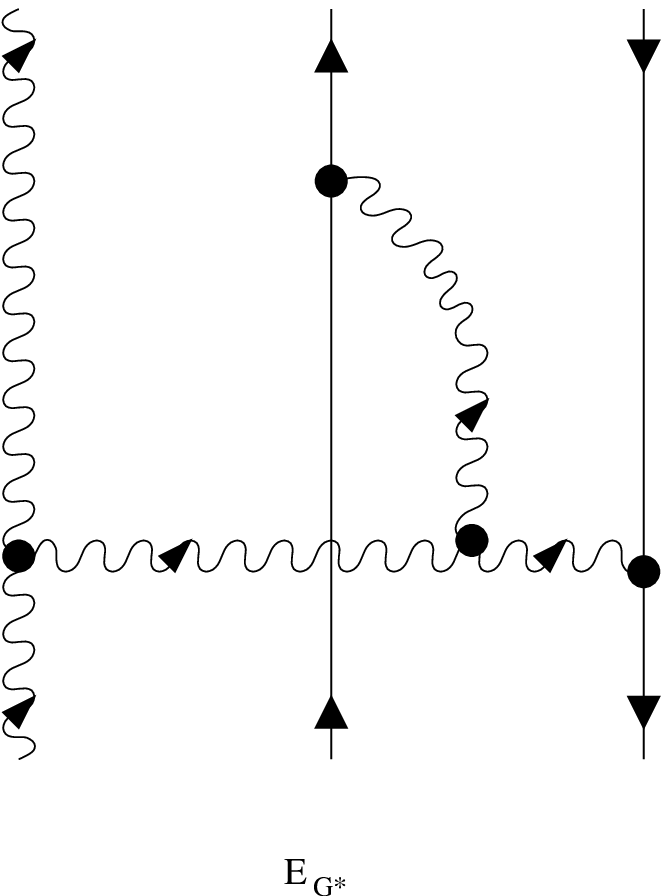}
      \vskip 26pt
    \includegraphics[width=42mm,height=65mm,angle=0]{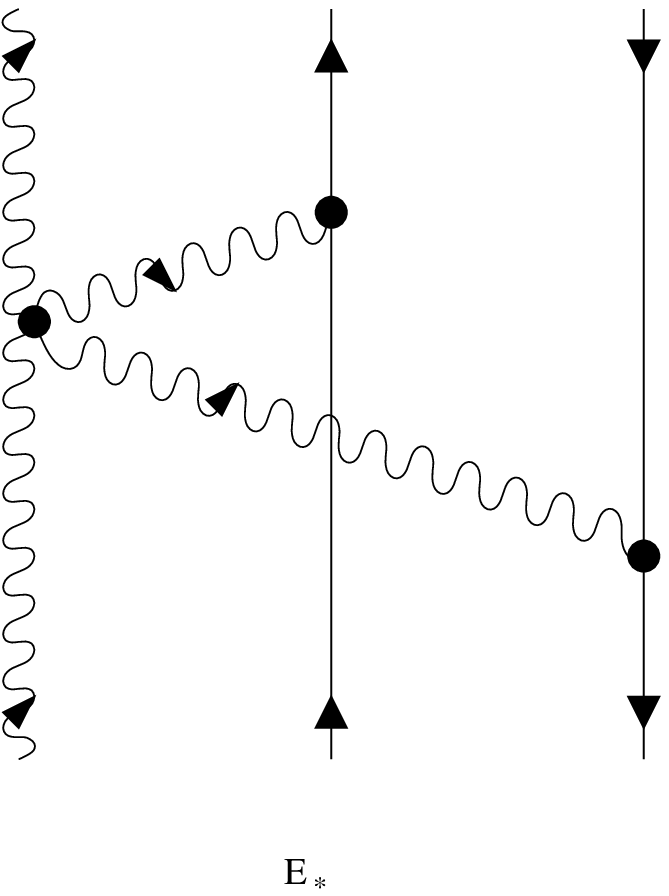}
      \hspace{1.2cm}
    \includegraphics[width=42mm,height=65mm,angle=0]{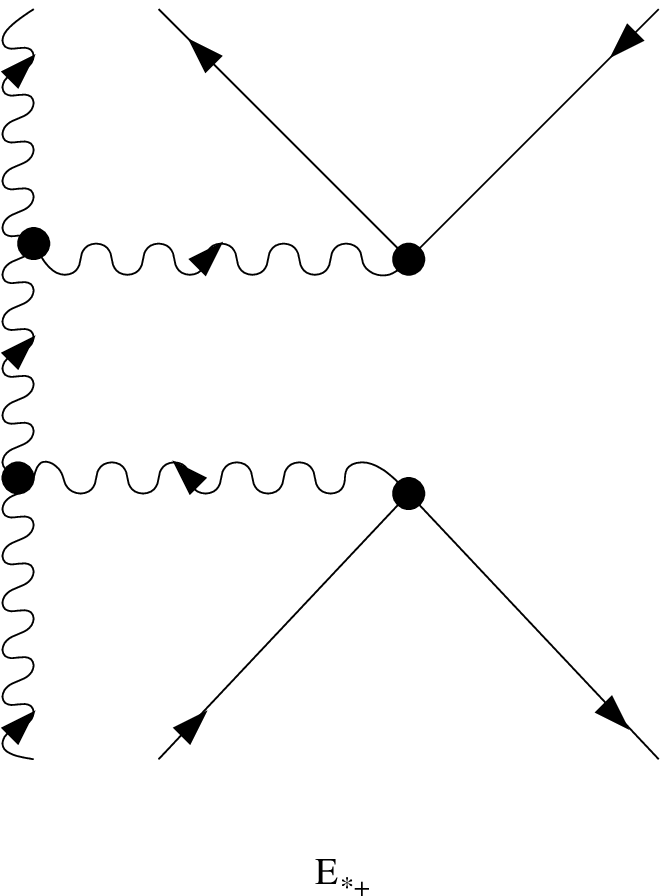}
      \hspace{1.2cm}
    \includegraphics[width=42mm,height=65mm,angle=0]{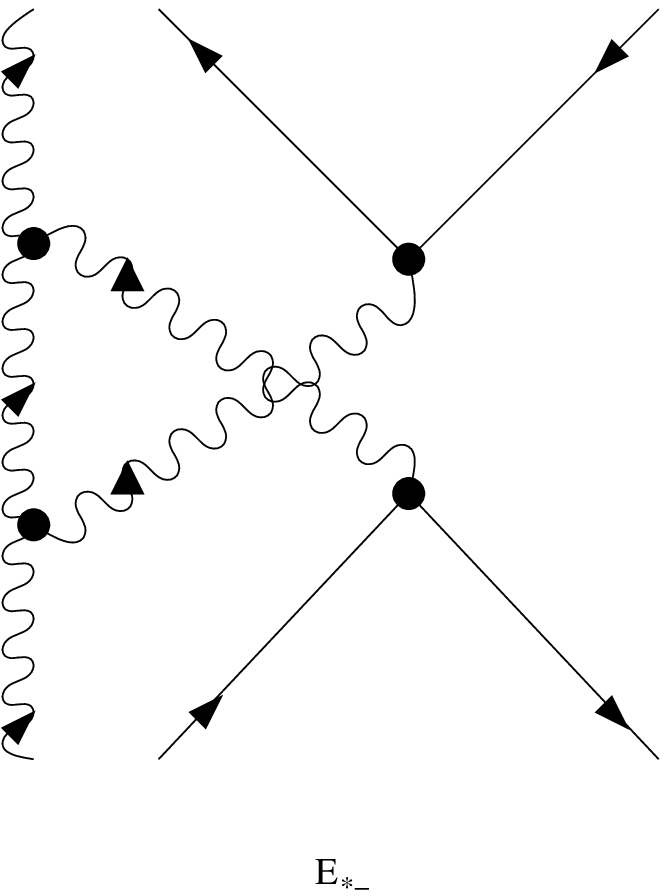}
      \vskip 26pt
    \includegraphics[width=42mm,height=65mm,angle=0]{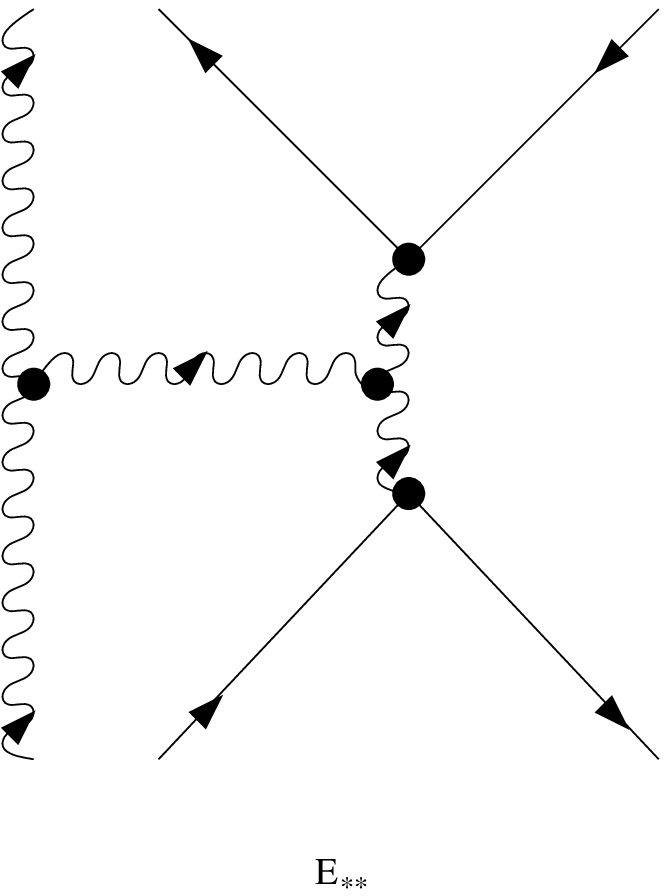}
      \hspace{1.2cm}
    \includegraphics[width=42mm,height=65mm,angle=0]{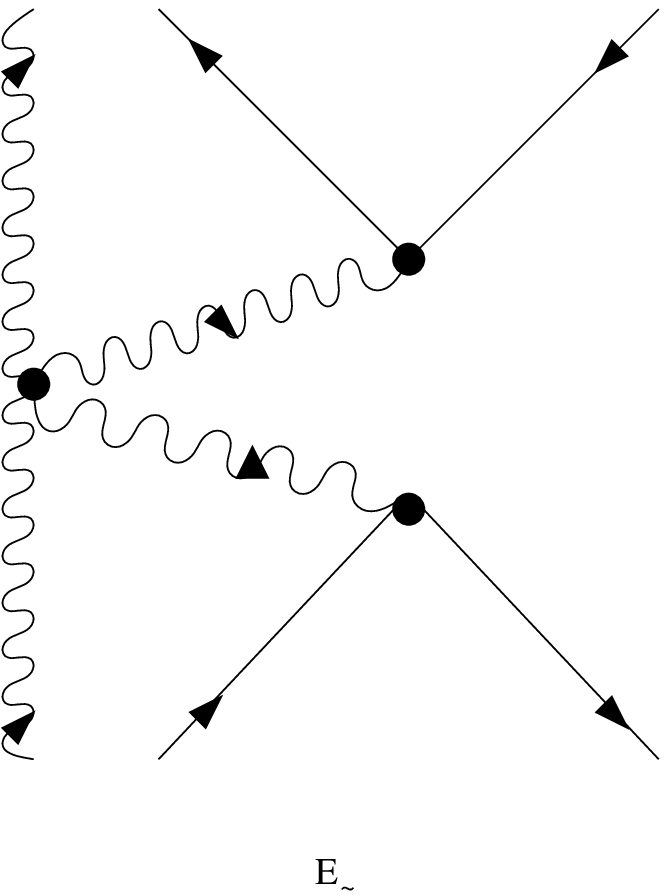}
\caption{Elastic gluon-quark-antiquark scattering.}
\label{fig12}
\end{figure}

\newpage
\begin{figure}
  \centering
    \includegraphics[width=42mm,height=65mm,angle=0]{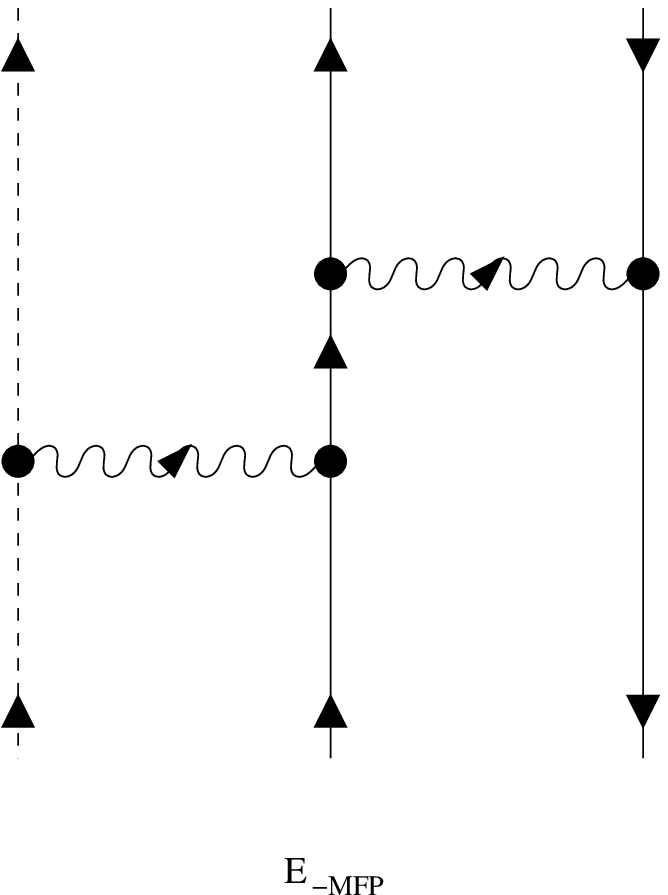}
      \hspace{1.2cm}
    \includegraphics[width=42mm,height=65mm,angle=0]{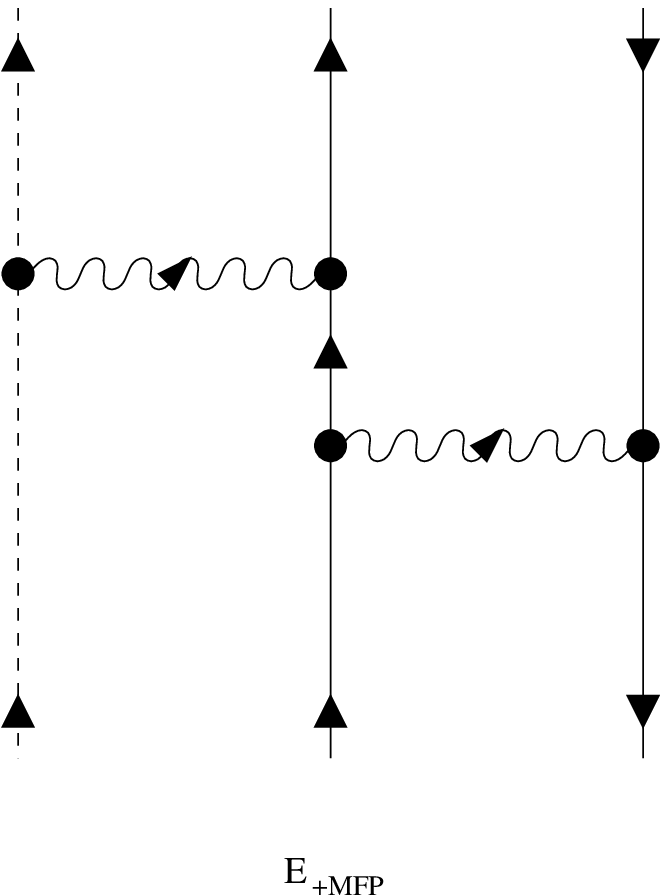}
      \hspace{1.2cm}
    \includegraphics[width=42mm,height=65mm,angle=0]{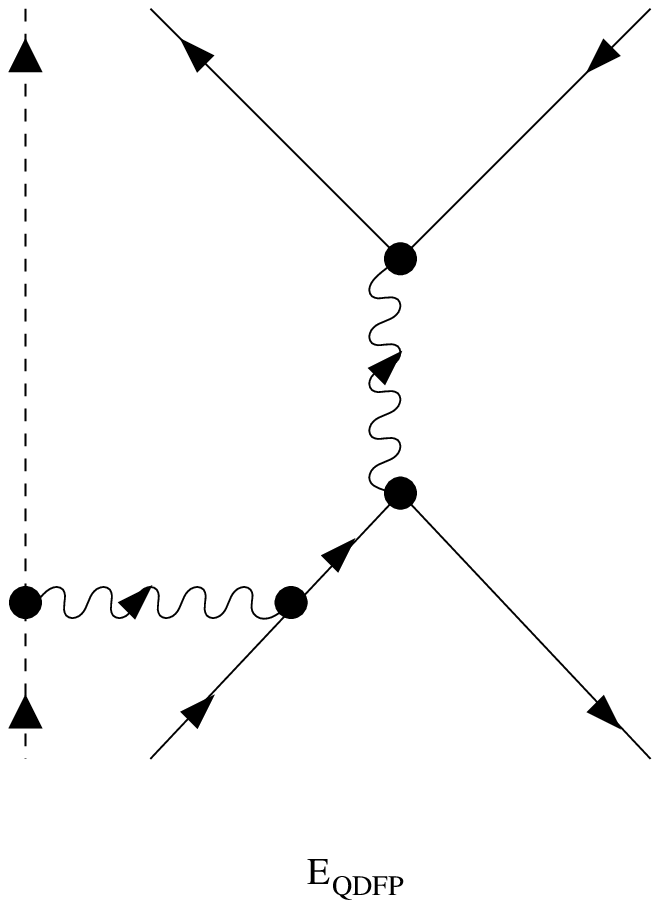}
       \vskip 26pt
    \includegraphics[width=42mm,height=65mm,angle=0]{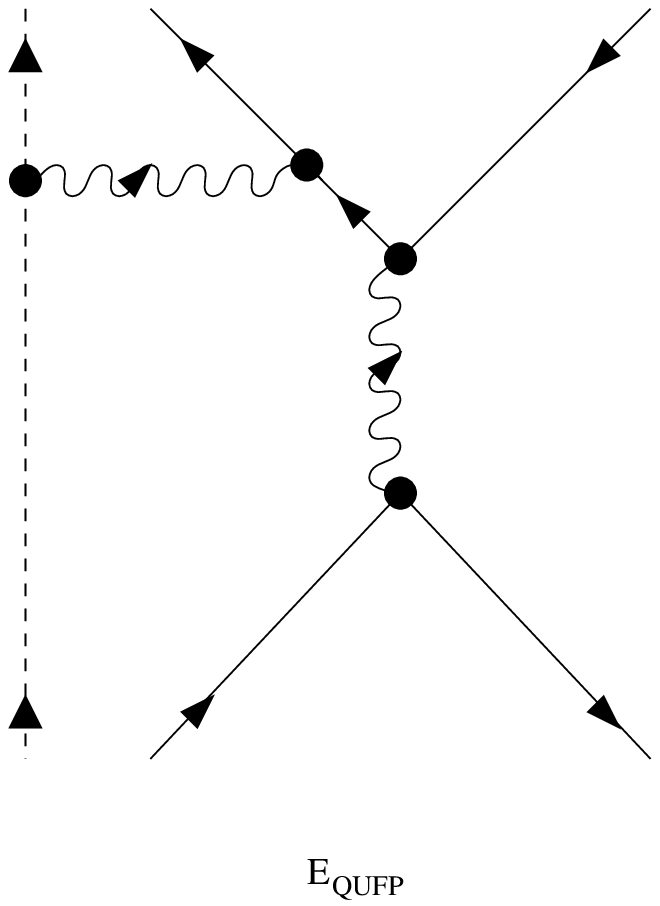}
      \hspace{1.2cm}
    \includegraphics[width=42mm,height=65mm,angle=0]{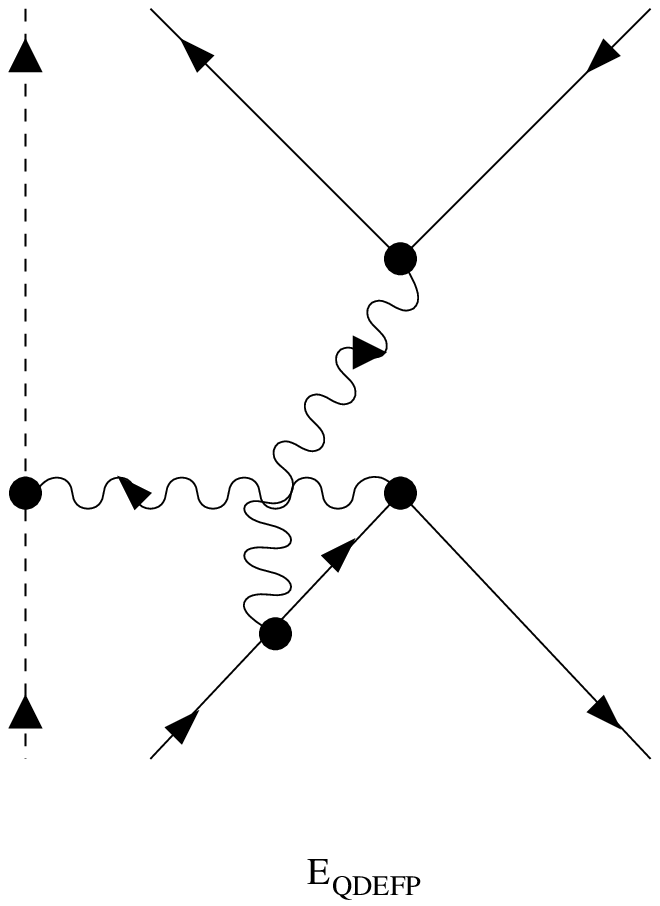}
      \hspace{1.2cm}
    \includegraphics[width=42mm,height=65mm,angle=0]{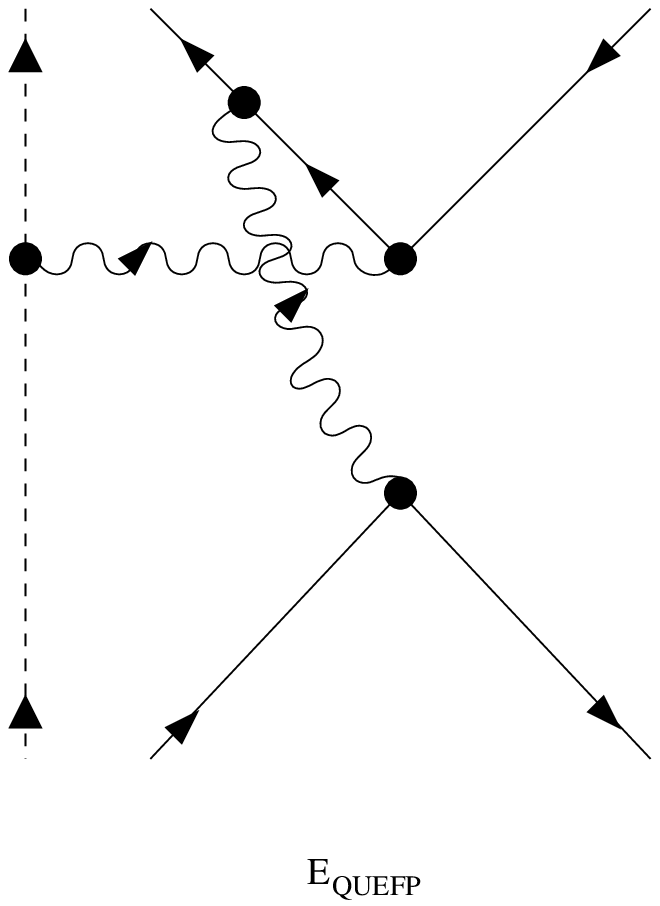}
\caption{Elastic ghost-quark-antiquark scattering.}
\label{fig13}
\end{figure}

\newpage
\begin{figure}
  \centering
    \includegraphics[width=42mm,height=65mm,angle=0]{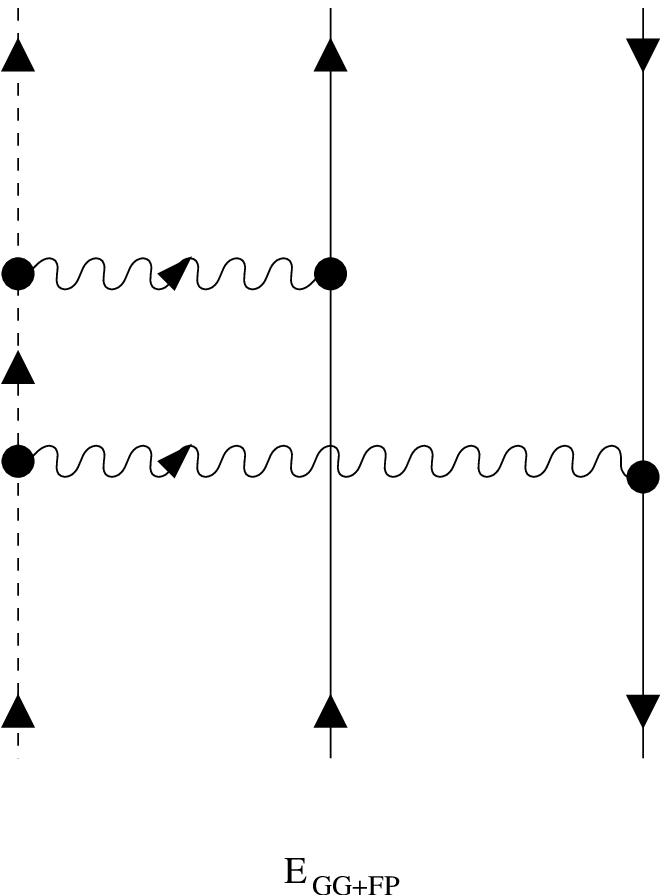}
      \hspace{1.2cm}
    \includegraphics[width=42mm,height=65mm,angle=0]{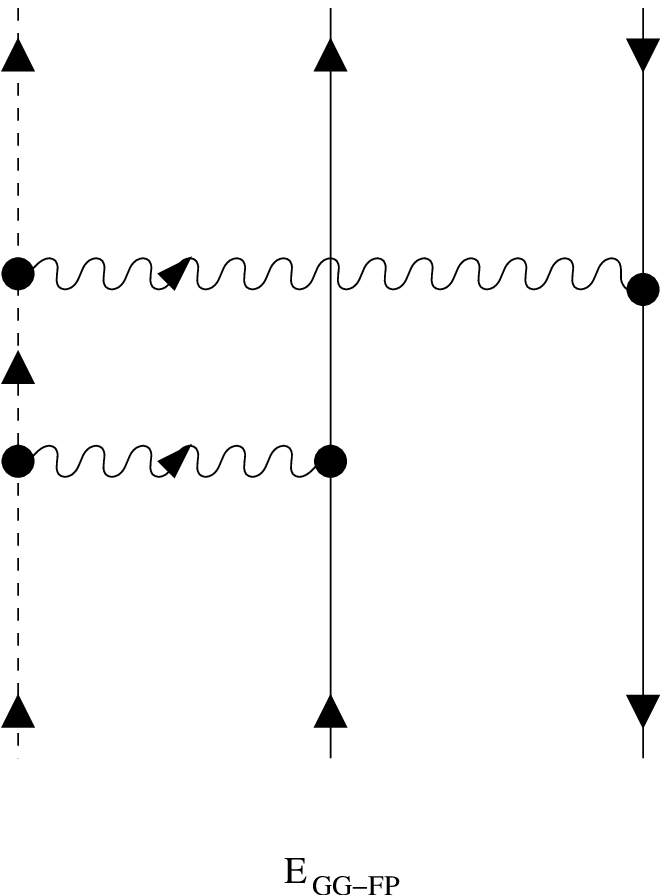}
      \hspace{1.2cm}
    \includegraphics[width=42mm,height=65mm,angle=0]{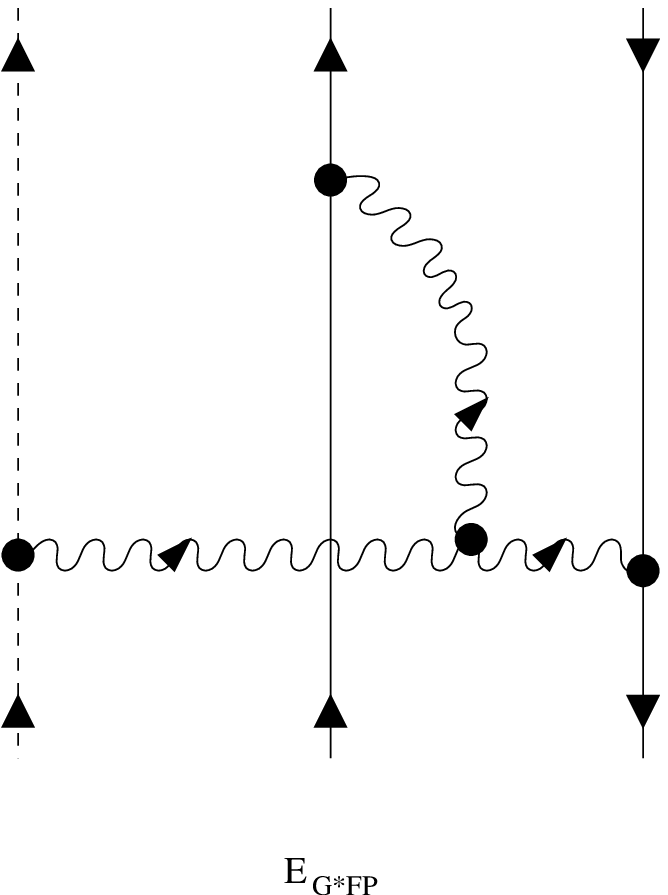}
       \vskip 26pt
    \includegraphics[width=42mm,height=65mm,angle=0]{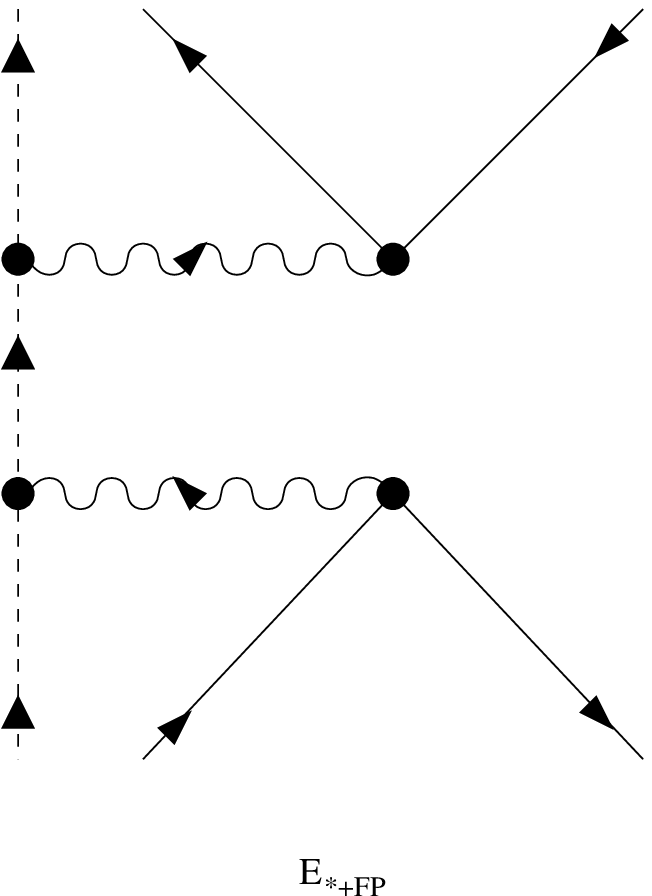}
      \hspace{1.2cm}
    \includegraphics[width=42mm,height=65mm,angle=0]{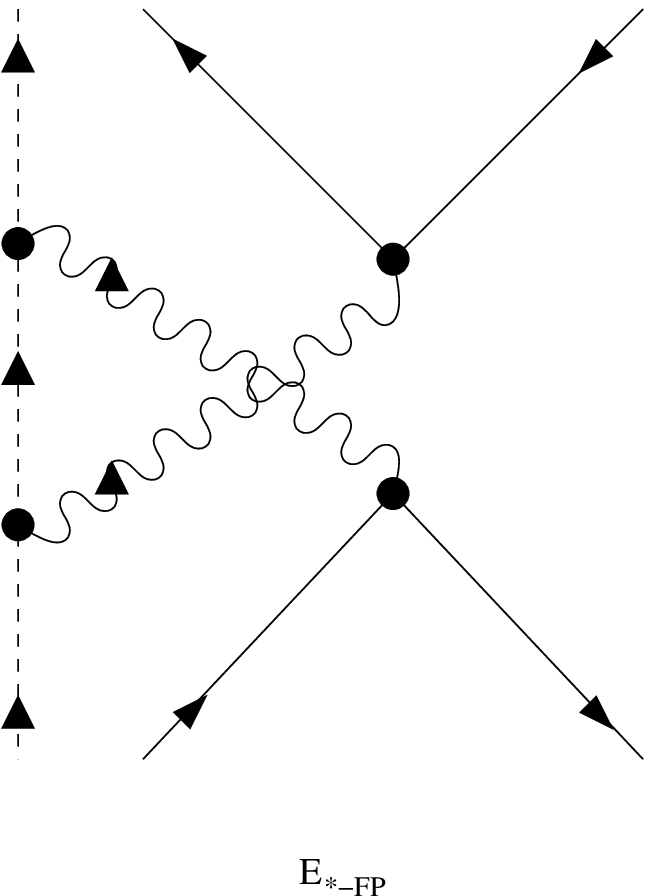}
      \hspace{1.2cm}
    \includegraphics[width=42mm,height=65mm,angle=0]{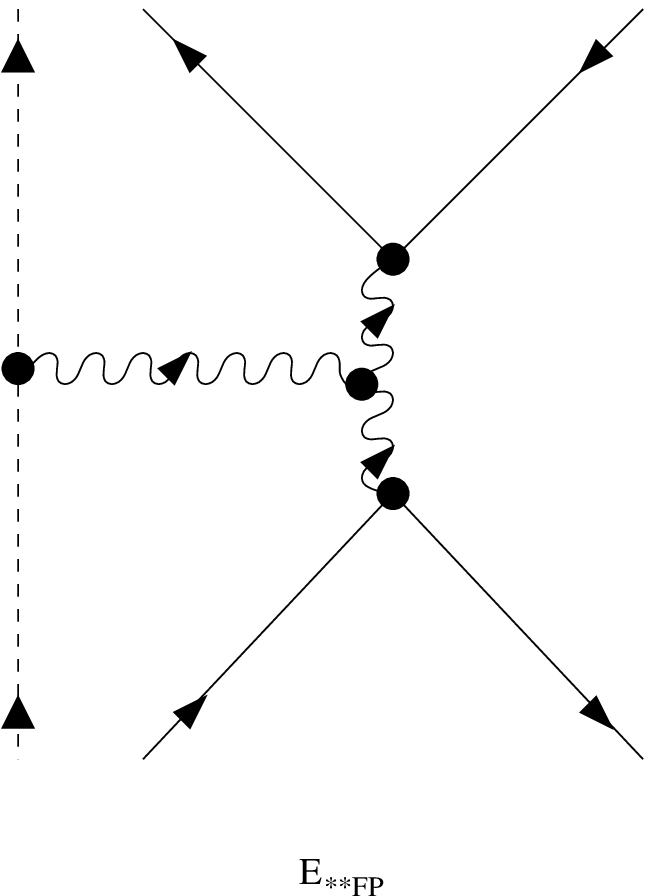}
\caption{Elastic ghost-quark-antiquark scattering.}
\label{fig14}
\end{figure}

\newpage
\begin{figure}
  \centering
    \includegraphics[width=120mm,height=80mm,angle=0]{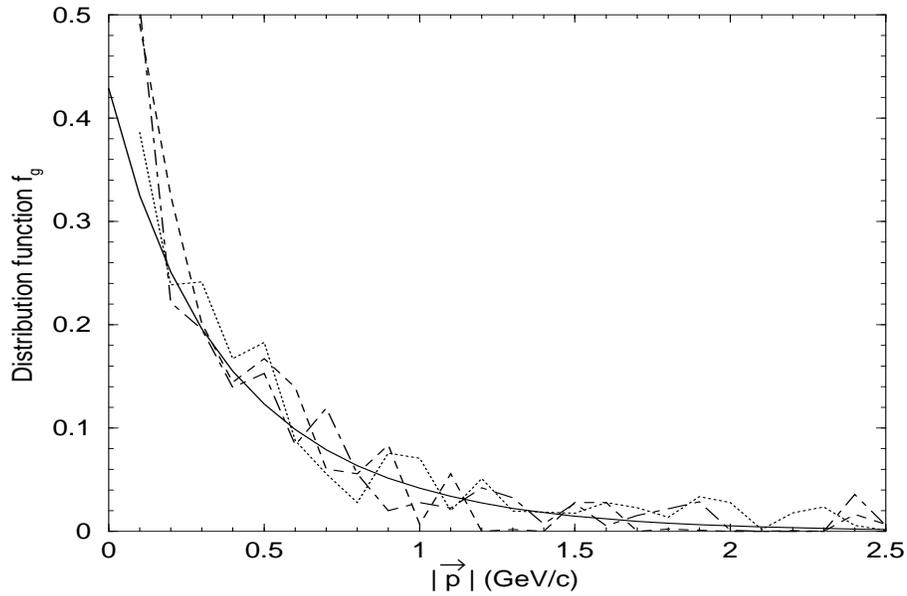}
\caption{Gluon distribution functions versus momentum in different directions
while gluon matter arrives at thermal equilibrium. The dotted, dashed and 
dot-dashed curves correspond to the angles relative to one incoming beam
direction $\theta =0^{\rm o}, 45^{\rm o}, 90^{\rm o}$, respectively.
The solid curve represents the thermal distribution function.}
\label{fig15}
\end{figure}

\newpage
\begin{figure}
  \centering
    \includegraphics[width=120mm,height=80mm,angle=0]{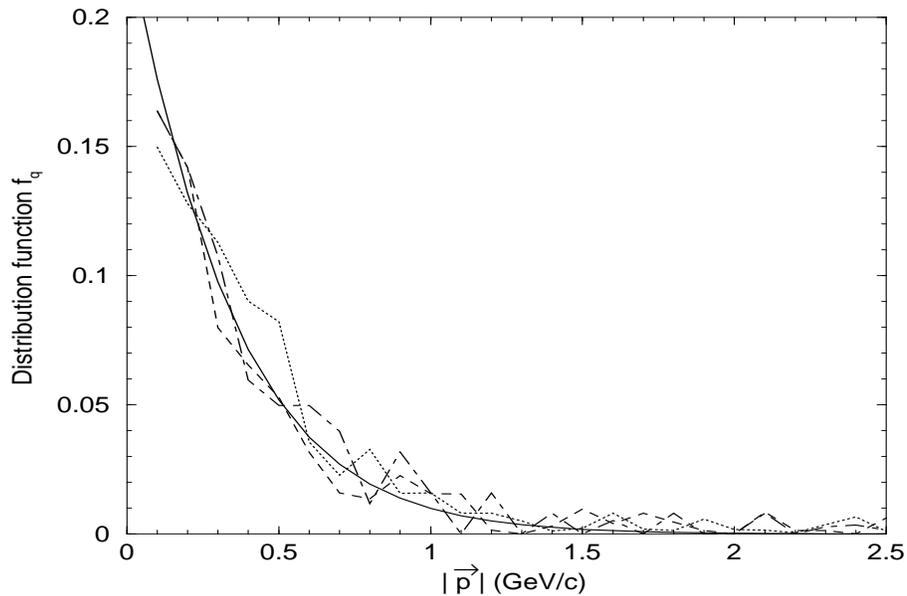}
\caption{The same as Fig. 15, except for quark distribution functions 
while quark matter arrives at thermal equilibrium.}
\label{fig16}
\end{figure}

\end{document}